\documentclass[aps,prb,reprint,twocolumn,superscriptaddress]{revtex4-2}

\usepackage{graphicx}
\usepackage{bm}
\usepackage{amsmath}  
\usepackage{amssymb}
\usepackage{hyperref}
\usepackage{xcolor}
\usepackage{relsize}
\usepackage{array}
\usepackage{multirow}
\usepackage{physics}
\usepackage{lipsum}
\usepackage[T1]{fontenc}
\usepackage{booktabs}

\renewcommand{\v}[1]{{\boldsymbol{#1}}}
\newcommand{\ov}{\overline}

\newcommand{\bq}{{\v q}}
\newcommand{\bk}{{\v k}}
\newcommand{\br}{{\v r}}
\newcommand{\bG}{{\v G}}

\renewcommand{\H}{\mathcal{H}}
\DeclareMathOperator{\ad}{ad}
\DeclareMathOperator{\diag}{diag}

\begin{document}

\title{Strong-coupling topological states and phase transitions in helical trilayer graphene} 

\author{Yves H. Kwan}
\affiliation{Princeton Center for Theoretical Science, Princeton University, Princeton NJ 08544, USA}
\author{Patrick J. Ledwith}
\affiliation{Department of Physics, Harvard University, Cambridge, MA 02138, USA}
\author{Chiu Fan Bowen Lo}
\affiliation{Department of Physics, Harvard University, Cambridge, MA 02138, USA}
\author{Trithep Devakul}
\affiliation{Department of Physics, Stanford University, Stanford, CA 94305, USA}
\affiliation{Department of Physics, Massachusetts Institute of Technology, Cambridge, MA 02139, USA}

\begin{abstract}
Magic-angle helical trilayer graphene relaxes into commensurate moir\'e domains, whose topological and well-isolated set of narrow bands possess ideal characteristics for realizing robust correlated topological phases, compared with other graphene-based moir\'e heterostructures. Combining strong-coupling analysis and Hartree-Fock calculations, we investigate the ground states at integer fillings $\nu$, and uncover a rich phase diagram of correlated insulators tuned by an external displacement field $D$. For small $D$, the system realizes several competing families of symmetry-broken generalized flavor ferromagnets, which exhibit various anomalous Hall signatures and Chern numbers as high as $|C|=6$. The interaction-induced dispersion renormalization is weak, so that the band flatness and the validity of strong-coupling theory are maintained at all integer fillings. For experimentally accessible displacement fields, the strong-coupling insulators at all $\nu$ undergo topological phase transitions, which appear continuous or weakly first-order. For larger $D$, we also find translation symmetry-broken phases such as Kekul\'e spiral order. Our results demonstrate the robust capability of helical trilayer graphene to host gate-tunable topological and symmetry-broken correlated phases, and lay the groundwork for future theoretical studies on other aspects such as fractional topological states.

\end{abstract}\maketitle

\section{Introduction}

\begin{figure}[t]
    \includegraphics[width=0.8\columnwidth]{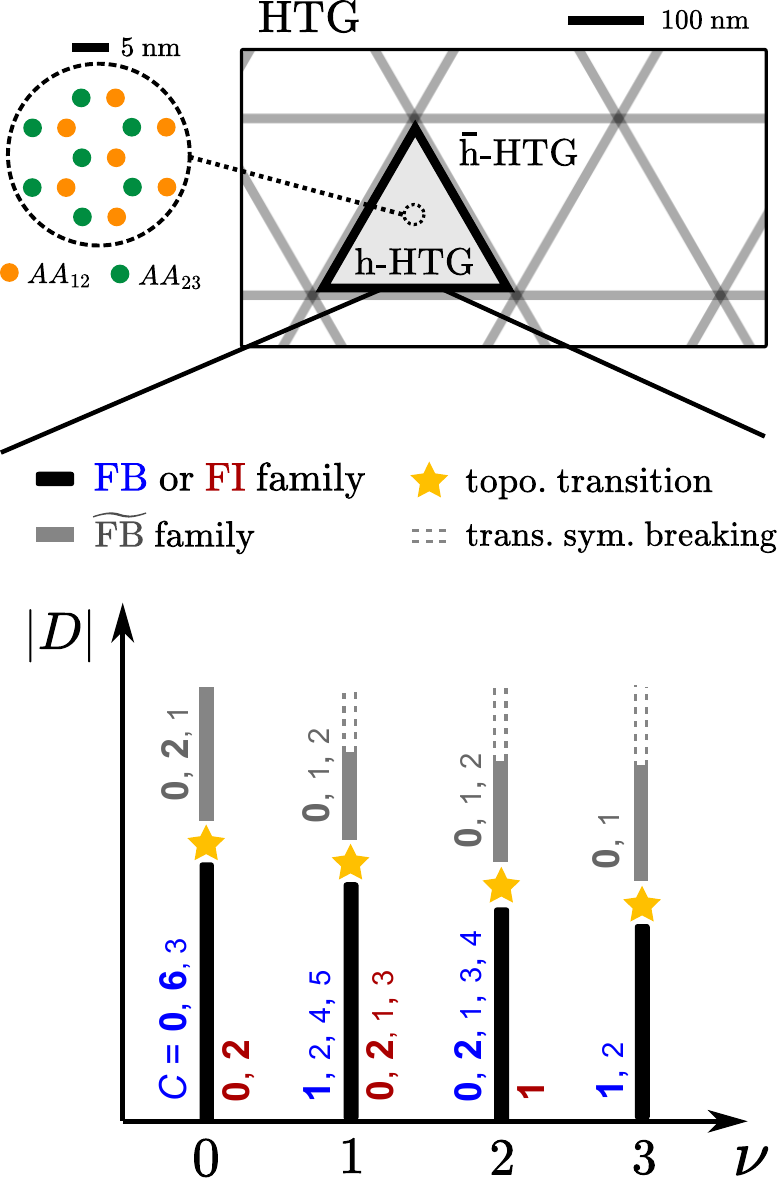}
    \caption{\textbf{Summary of correlated insulating phases in h-HTG at integer fillings $\nu$ as a function of displacement field $D$.} \textbf{Top:}  The super-moir\'e structure of helical trilayer graphene (HTG) relaxes into large moir\'e-periodic domains of h-HTG and $\bar{\text{h}}$-HTG separated by gapless domain walls (grey). h-HTG realizes a honeycomb lattice consisting of $AA$ stacking regions of the two pairs of adjacent layers. \textbf{Bottom:} Schematic phase diagram of h-HTG. At small $D$, there are two competing strong-coupling families of low-lying states, flavor-balanced (FB, blue) and flavor-imbalanced (FI, red), defined in Sec.~\ref{sec:strong_coupling}. They are characterized by a preference towards partial or full occupation of flavors respectively. The possible Chern numbers $|C|$ are shown, with large bold entries corresponding to the primary ground states predicted by our analysis. At a displacement field-tuned topological transition (stars), the system enters the $\widetilde{\text{FB}}$ phase whose sets of Chern numbers differs from the zero-field case. At even higher fields, the system further realizes various translation symmetry-breaking phases (dotted lines).}
    \label{fig:D_nu_splash}
\end{figure}

The coexistence of narrow bands, electronic topology and strong interactions provides a fertile ground for realizing fascinating quantum phases of matter. A now-classic example is realized by the fractional quantum Hall effect (FQHE) in 2D semiconductor quantum wells~\cite{tsuiFQHE1982}, where the requisite conditions are generated by the Landau quantization induced by an external magnetic field. The presence of flavor degrees of freedom, like spin, valley, layer or orbital components, further enriches the physics, even at integer fillings, where spontaneous symmetry-breaking introduces new collective phenomena such as topological defects and additional quantized responses~\cite{girvin2002quantum}. Various platforms have been shown to experimentally realize such quantum Hall ferromagnetism (QHFM), such as quantum Hall bilayers~\cite{eisenstein2014exciton} and the zeroth Landau level (LL) of graphene~\cite{goerbigrmp2011}. These systems are also typically associated with an enlarged manifold of nearly-degenerate orders.

Magic-angle twisted bilayer graphene (TBG)~\cite{cao2018correlated,cao2018unconventional,yankowitz2019tuning,lu2019superconductors} has attracted remarkable attention as the poster child of the family of correlated moir\'e materials~\cite{andrei2021marvels,mak2022semiconductor}, and has accumulated an ever-growing catalog of experimentally-observed phenomena. Theoretically, this system holds promise for possessing all of the above ingredients without the need for an external magnetic field. The spatially-modulated interlayer tunneling is responsible for the small dispersion, the Dirac points imbue the moir\'e bands with non-trivial topology, and the graphene layers supply the valley, spin, and sublattice degrees of freedom. This notion of interactions dominating a set of narrow topological bands in TBG is formalized in the ``strong-coupling'' framework~\cite{bultinckGroundStateHidden2020,lianTBGIVExact2020,ledwithStrongCouplingTheory2021}, which enables a controlled analysis of various deviations from an idealized solvable limit with completely flat bands and enhanced symmetries~\cite{tarnopolskyOriginMagicAngles2019}. The result is a manifold of closely-competing symmetry-broken correlated insulating states, akin to generalized QHFM. However, while predicted to arise at various fillings in the strong-coupling limit, such topological states are often overpowered by competing non-topological states in TBG under realistic conditions. Part of the reason is due to the large interaction-induced ``Hartree'' dispersion at finite density~\cite{Guineaelectrostatic2018,ardemakercharge2019,ceapinning2019,goodwin2020hartree,Kangcascades2021,pierce2021unconventional,parkerFieldtunedZerofieldFractional2021}, which originates from the real-space inhomogeneity of the moir\'e wavefunctions. This drives the system away from the strong-coupling regime, especially in the presence of strain~\cite{kerelsky2019maximized,choi2019electronic,xieSpectroscopicSignaturesManybody2019,mespleheterostrain2021,nuckolls2023quantum,parkerStrainInducedQuantumPhase2021,kwan2021kekule,wagner2022global,wang2022kekul}, with the consequence being that an applied magnetic field or substrate alignment is often necessary to stabilize such topological states~\cite{serlinIntrinsicQuantizedAnomalous2020,sharpeEmergentFerromagnetismThreequarters2019,stepanovCompetingZeroFieldChern2021,nuckollsStronglyCorrelatedChern2020,wuBlochModelWave2013,dasSymmetrybrokenChernInsulators2021,saitoHofstadterSubbandFerromagnetism2021,xieFractionalChernInsulators2021,parkerFieldtunedZerofieldFractional2021,bultinckMechanismAnomalousHall2020,zhangNearlyFlatChern2019}

Recently, correlated topological states have been proposed to arise in helical trilayer graphene (HTG), 
a structure consisting of three graphene layers with identical twist angles $\theta$ between adjacent layers~\cite{devakul2023magicangle} (see also Refs.~\cite{mora2019flatbands,zhu2020twisted,mao2023supermoire,popov2023magic,guerci2023chern,nakatsuji2023multiscale,popov2023butterfly,foo2023extended,guerci2023nature}). 
In the absence of lattice relaxation, the pairs of adjacent layers in HTG form two moir\'e lattices, which themselves form a super-moir\'e lattice at very long lengthscales ($a_{mm}\simeq 250\,\text{nm}$ near $\theta\simeq 1.8^\circ$).
Theoretical analysis demonstrated that lattice relaxation plays a key role and leads to the formation of a commensurate single-moir\'e structure (with moir\'e periodicity $a_m\simeq 8\,\text{nm}$) over large regions, made possible by the slight elastic deformation of the graphene layers~\cite{devakul2023magicangle,nakatsuji2023multiscale}. These commensurate regions come in two ${C}_{2z}$-related versions, called h-HTG and $\bar{\text{h}}$-HTG, which are tiled together and separated by a triangular network of narrow gapless domain walls (see Fig.~\ref{fig:D_nu_splash}).
In the resulting structure of h-HTG, the AA regions of the two moir\'e lattices come together to form a moir\'e-scale honeycomb configuration.  Henceforth, we focus on the physics of commensurate h-HTG, though our results straightforwardly generalize to $\bar{\text{h}}$-HTG as well.

The most remarkable aspect of h-HTG lies in its electronic properties.  The single particle electronic structure in each spin and valley sector consists of a pair of flat topological bands with Chern numbers $|C|=1,2$ and near-ideal quantum geometry. Because the relaxed structure breaks $C_{2z}$ symmetry, the flat bands carry non-zero total valley-Chern number.  
Furthermore, the flat-band manifold is isolated from remote bands by a significant energy gap $E_{\mathrm{rem.~gap}}\simeq 100\,$meV, much larger than the interaction energy scale. 
These features suggest that h-HTG is an ideal platform for exploring interaction-dominated physics in topological bands, 
with potential for realizing exotic topological states such as integer and fractional Chern insulators at zero magnetic field.
These findings call for a detailed theoretical study of the interaction-driven physics.

In this work, we perform a comprehensive analysis of the interacting phase diagram of h-HTG at integer fillings.
We employ self-consistent Hartree-Fock mean field theory and strong coupling analysis, which reveals a rich phase structure that is highly dependent on an externally applied displacement field. For weak displacement fields, we uncover a plethora of closely-competing symmetry-broken topological states with Chern numbers as high as $|C|=6$ (see Fig.~\ref{fig:D_nu_splash}), which are well captured within strong-coupling perturbation theory. We find that, compared to other graphene-based moir\'e systems, the Hartree corrections are weak owing to the relatively homogeneous charge density of the central-band wavefunctions. This maintains the stability of strong-coupling correlated insulators at non-zero integer fillings, safeguards against mixing with remote bands, and allows for relatively flat quasiparticle bands even when accounting for interaction renormalization. 

For critical displacement fields well within experimental capabilities, our calculations show that all integer filling factors can undergo continuous or weakly first-order topological phase transitions to states with smaller or vanishing Chern numbers (see schematic of Fig.~\ref{fig:D_nu_splash}). Interestingly, the states just above the transition still preserve significant strong-coupling character, in that they retain sizable flavor and/or sublattice polarizations close to the zero-field case. This is possible because the transition involves a band inversion that is localized in momentum space. Hence, h-HTG potentially realizes the universal theory of Dirac mass inversion in the strongly-interacting regime, away from weak-coupling where it is normally studied. Importantly, such physics is experimentally accessible by tuning the displacement field.

For yet larger fields, we find translation-symmetry-breaking phases such as charge density waves, as well as a Kekul\'e spiral order~\cite{kwan2021kekule,wagner2022global,wang2022kekul} that has connections to that recently imaged in TBG~\cite{nuckolls2023quantum} and mirror-symmetric trilayer graphene~\cite{kim2023imaging} (which differs from HTG in that it has alternating twists between the layers).

Our results highlight HTG as a highly-tunable system that exhibits a panoply of orbital Chern insulators, symmetry-breaking orders, and displacement-tuned topological transitions. The phenomena uncovered in this paper can be studied experimentally through various probes. From a theoretical standpoint, we argue that h-HTG presents a near-ideal moir\'e platform for quantum-Hall-like strong coupling physics, including more exotic phases like fractional Chern insulators~\cite{parameswaranFractionalQuantumHall2013,bergholtzTOPOLOGICALFLATBAND2013,liuRecentDevelopmentsFractional2022,neupertFractionalQuantumHall2011,shengFractionalQuantumHall2011,regnaultFractionalChernInsulator2011,scaffidiAdiabaticContinuationFractional2012,royBandGeometryFractional2014,kourtisFractionalChernInsulators2014}, and is relatively free from the various complications that are present in other related systems. We also discuss various extensions such as non-integer fillings, and the impact of the super-moir\'e structure of domains in HTG.

\section{Model and methods}

\begin{figure}[t]
    \includegraphics[width=1\columnwidth]{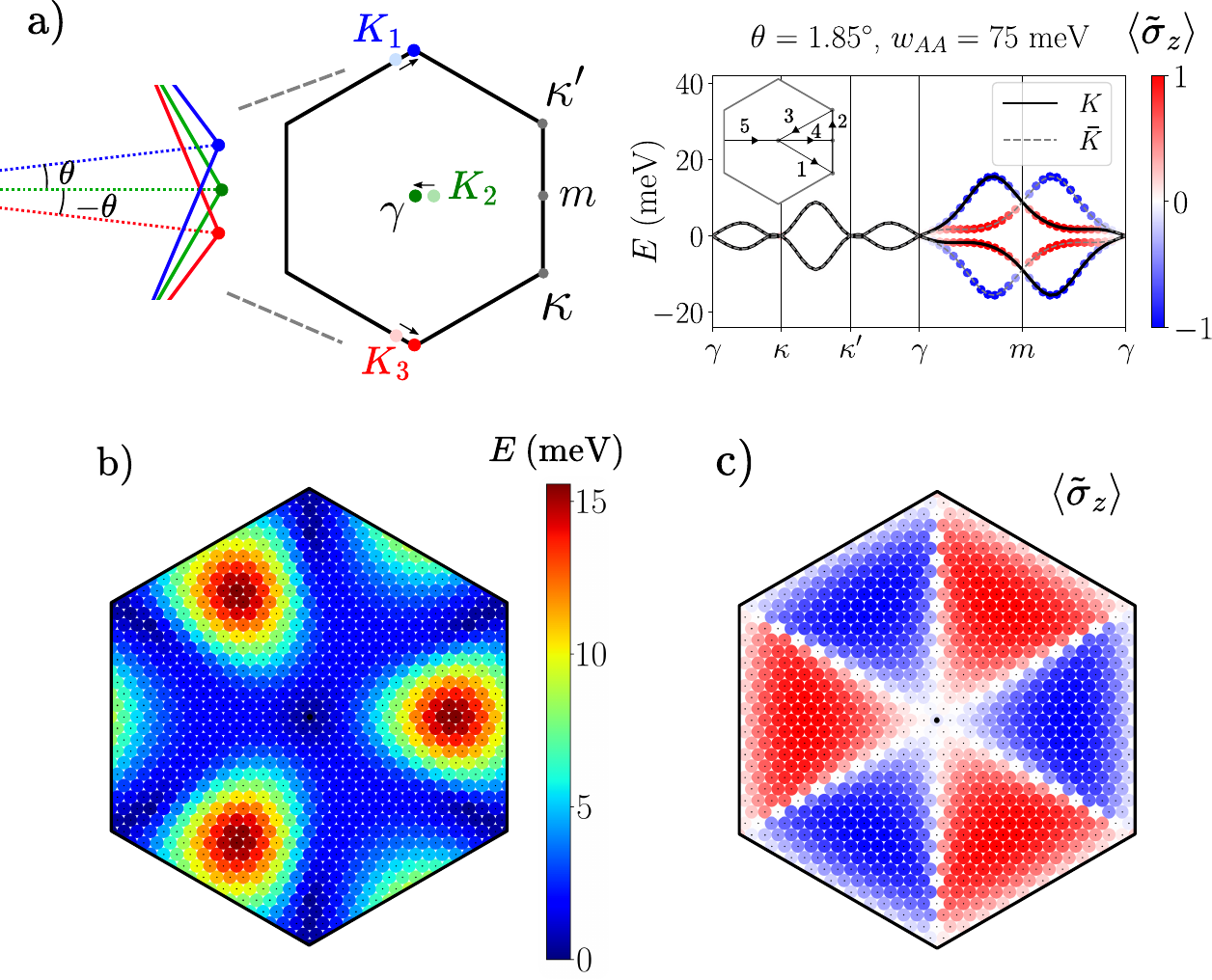}
    \caption{\textbf{Non-interacting band structure at zero displacement field.} a) Left: The moir\'e Brillouin zone (mBZ) is determined from the Dirac momenta of the three layers, which relax slightly (arrows) to form a locally commensurate structure. Right: Dispersion of the central bands along a path in the mBZ shown in the inset. Color indicates polarization in the Chern-sublattice basis $\langle\tilde{\sigma}_z\rangle$, where $\tilde{\sigma}_z=+1 (-1)$ is sublattice $A$ ($B$). b,c) Energy dispersion and Chern-sublattice polarization of the $K$-valley conduction band in the mBZ. }
    \label{fig:BM_zeroU}
\end{figure}

\subsection{Non-interacting continuum model}

Our starting point for studying h-HTG is a generalization of the Bistritzer-MacDonald (BM) continuum model for valley $K$ ($\tau=+$)~\cite{bistritzerMoireBandsTwisted2011,devakul2023magicangle}
\begin{equation}\label{eq:nonint_BM}
H^\text{BM}_K=\begin{bmatrix}
    -iv_F\bm{\sigma}\cdot\nabla & T(\bm{r}-\bm{d}_t) & 0 \\
    T^\dagger(\bm{r}-\bm{d}_t) & -iv_F\bm{\sigma}\cdot\nabla & T(\bm{r}-\bm{d}_b)\\
    0 & T^\dagger(\bm{r}-\bm{d}_b) & -iv_F\bm{\sigma}\cdot\nabla
\end{bmatrix}
\end{equation}
where the matrix acts on layer space $l=1,2,3$, $\bm{\sigma}=(\sigma_x,\sigma_y)$ acts on the microscopic sublattice, and the graphene Dirac velocity $v_F=8.8\times 10^5\,\text{ms}^{-1}$~\cite{bistritzerMoireBandsTwisted2011}. The Hamiltonian for valley $\bar{K}$ ($\tau=-$) can be found by time-reversal. Combined with spin $s=\uparrow,\downarrow$, there are four spin-valley flavors. Note that Eq.~\ref{eq:nonint_BM} has been written in a layer-boosted frame such that a Bloch function at momentum $\bm{k}$ satisfies $\psi_{\bm{k},l}(\bm{r}+\bm{a}_i)=e^{i(\bm{k}-\bm{K}_l)\cdot \bm{a}_i}\psi_{\bm{k},l}(\bm{r})$, where $\bm{K}_l$ is the layer-dependent Dirac momentum, suitably deformed to allow for a commensurate structure. $\bm{K}_2$ folds onto the moir\'e $\gamma$ point, while $\bm{K}_1$ ($\bm{K}_3$) folds onto $\kappa$ ($\kappa'$) (Fig.~\ref{fig:BM_zeroU}a). $\bm{a}_i$ is a basis moir\'e lattice vector $\bm{a}_{1,2}=\frac{4\pi}{3k_\theta}(\pm\frac{\sqrt{3}}{2},\frac{1}{2})$, where $k_\theta=2K_D\sin\frac{\theta}{2}$, with $K_D$ the Dirac wavevector.

The interlayer tunneling takes the form
\begin{equation}\label{eq:nonint_BM_tunnel}
\begin{gathered}
    T(\bm{r})=\begin{bmatrix}
        w_{AA}t_0(\bm{r}) & w_{AB}t_{-1}(\bm{r})\\
        w_{AB}t_1(\bm{r}) & w_{AA}t_0(\bm{r})
    \end{bmatrix}\\
    t_\alpha(\bm{r})=\sum_{n=0}^{2}e^{\frac{2\pi i}{3}n\alpha}e^{-i\bm{q}_n\cdot\bm{r}}\\
    q_{n,x}+iq_{n,y}=-ik_\theta e^{\frac{2\pi i}{3}n},
\end{gathered}
\end{equation}
where $\bm{K}_{1,3}=\mp \bm{q}_0+\bm{K}_2$. Lattice relaxation~\cite{namLatticeRelaxationEnergy2017,carr2019exact} and renormalization~\cite{vafekRenormalizationGroupStudy2020} effects lead to a suppression of the chiral ratio $\kappa=\frac{w_{AA}}{w_{AB}}<1$, whose precise value is difficult to pin down. We fix $w_{AB}=110\,\text{meV}$~\cite{bistritzerMoireBandsTwisted2011}, but allow $w_{AA}$ to vary. For most calculations, we set the chiral ratio to a physically reasonable value $\kappa\simeq 0.7$~\cite{namLatticeRelaxationEnergy2017,carr2018relaxation,carr2019exact,guinea2019continuum,ledwithTBNotTB2021,carrMinimalModelLowenergy2019,koshinoEffectiveContinuumModel2020,koshinoMaximallyLocalizedWannier2018,dasSymmetrybrokenChernInsulators2021,vafekRenormalizationGroupStudy2020}. The relative interlayer moir\'e shift corresponding to the structure of h-HTG is given by $\bm{d}_t-\bm{d}_b=\bm{\delta}=\frac{1}{3}(\bm{a}_2-\bm{a}_1)$.  

To model the effect of a displacement field $D$, which is tunable in dual-gated samples, we add an interlayer potential $U$ to Eq.~\ref{eq:nonint_BM}, such that layers $l=1,2,3$ have energy shifts $+U,0,-U$. The relation between the two is $U=d_\text{inter}D/\epsilon_\perp$, where $d_\text{inter}=3.3\,\text{\AA}$ is the interlayer distance and $\epsilon_\perp$ is the perpendicular dielectric constant (up to electrostatic corrections which must be taken into account self-consistently~\cite{uri2023superconductivity}). The largest displacement fields attainable in experiment around charge neutrality are $D/\epsilon_0\sim 1\,\text{V/nm}$, which corresponds to $U\sim 80\,\text{meV}$ assuming $\epsilon_\perp\simeq 4$. We only consider positive $U$ since negative values are related by symmetry.

As shown in Fig.~\ref{fig:BM_zeroU}a,b for $U=0$, the central non-interacting BM bands (two per flavor) become narrow with bandwidth $W\sim 20\,\text{meV}$ near the magic angle $\theta\sim 1.8^\circ$. Since the remote band gaps are large $\sim 100\,\text{meV}$, we project the system into the central bands. The topology of the bands is manifest in the Chern basis (also called the sublattice basis), obtained by diagonalizing the microscopic sublattice operator $\sigma_z$ within this subspace~\cite{bultinckGroundStateHidden2020,lianTBGIVExact2020,khalafChargedSkyrmionsTopological2021,ledwithStrongCouplingTheory2021}. 
We introduce a Chern-sublattice label $\tilde{\sigma}_z=A,B$ according to the predominant microsopic sublattice polarization. Unless otherwise stated, the term `sublattice' refers directly to this label rather than the microsopic sublattice polarization. The Chern bands can be therefore indexed by $(\tau,s,\tilde{\sigma})$, with Chern numbers~\cite{devakul2023magicangle}
\begin{equation}\label{eq:Chern_numbers}
\begin{gathered}
    C_{K,s,A}=1,\quad C_{K,s,B}=-2\\
    C_{\bar{K},s,A}=-1,\quad C_{\bar{K},s,B}=2.
\end{gathered}
\end{equation}
The most dispersive parts of the BM bands, which lie along the $\gamma-m$ lines in the moir\'e Brillouin zone (mBZ), predominantly arise from the $B$ sublattice (Fig.~\ref{fig:BM_zeroU}b,c). Meanwhile, the kinetic energy of Bloch states dominated by the $A$ sublattice remains small. Note that the microscopic sublattice polarization of the $A$-bands is less than that of the $B$-bands, which is allowed due to the lack of any symmetry that interchanges sublattice.

\subsection{Band-projected interacting model}
The continuum model is augmented with long-range dual-gate screened Coulomb interactions $V(q)=\frac{e^2}{2\epsilon_0\epsilon_r q}\tanh qd_{\text{sc}}$, where the relative permittivity $\epsilon_r=8$ captures the effect of the hBN dielectric and remote bands, and the screening length $d_{\text{sc}}=25\,\text{nm}$.  A subtraction scheme is required to prevent double-counting interactions, as they already feed into model parameters such as the Fermi velocity~\cite{xieNatureCorrelatedInsulator2020,bultinckGroundStateHidden2020,parkerFieldtunedZerofieldFractional2021,kwan2021kekule,kwanskyrmions2022}. We will use the `average' scheme where the electron density is measured with respect to a reference density corresponding to half-filling of each flat band at infinite temperature~\cite{lianTBGIVExact2020,parkerFieldtunedZerofieldFractional2021}. We neglect terms such as intervalley Coulomb and phonon-induced contributions which scatter electrons between the valleys and are suppressed~\cite{chatterjeeSymmetryBreakingSkyrmionic2020}. The full band-projected Hamiltonian is
\begin{equation}
\H = \frac{1}{2A} \sum_{\bq \in \mathbb{R}^2} V(\bq) \delta \hat{\rho}_\bq \delta \hat{\rho}_{-\bq} + \sum_{\bk \in \text{mBZ}} c^\dag_\bk h(\bk) c_\bk,
\label{eq:intham_main}
\end{equation}
where $A=N_MA_M$ is the total area, $c^\dagger_\bk$ is a moir\'e band creation operator with spinor structure in flavor and band space, and $\delta \hat{\rho}_\bq = \hat{\rho}_{\bq} - 4 \ov{\rho}_\bq$ is the density measured with respect to half-filling of all bands (there are eight bands in total, and we define $\ov{\rho}_\bq$ to be the average band density). We note that while that $\bk$ varies over the mBZ, and labels the states of various Bloch momenta, the momentum $\bq$ varies over the entire plane because density fluctuations have a continuous profile and vary within a single moir\'e unit cell.

The form of the density operators are $\hat{\rho}_\bq = \sum_\bk c^\dag_\bk \Lambda_\bq(\bk) c_{\bk + \bq}$ and $\overline{\rho}_\bq = \frac{1}{2}(\ov{\rho}_\bq^{A} + \ov{\rho}_\bq^B)$, where $\overline{\rho}^{\tilde\sigma}_{\bq} = \sum_\bG \delta_{\bq,\bG} \sum_{\bk} \Lambda^{\tau, \tilde{\sigma} \tilde{\sigma} }_\bG(\bk)$ is the translationally symmetric background density in a periodic gauge $c_\bk = c_{\bk + \bG}$.  
The density operators are written in terms of the form factor matrix $\Lambda^{\tau \tilde{\sigma} \tilde{\sigma}'}_\bq(\bk) = \langle u^\tau_{\bk \tilde{\sigma}}| u^\tau_{\bk + \bq \tilde{\sigma}'}\rangle$ which consists of the amplitudes for the density operator to scatter Bloch states, $\ket{u^\tau_{\bk \tilde{\sigma}}}$, at wavevector $\bk + \bq$ to $\bk$. Note that while the form factors depend on $\tau$, the densities do not due to time-reversal symmetry. Since we are in the sublattice basis, $h(\bk)$ is a matrix that is block diagonal in valley and spin but has both diagonal and off-diagonal matrix elements in sublattice $\tilde{\sigma}$. 

The interacting Hamiltonian $\H$ has a large set of symmetries~\cite{devakul2023magicangle}. On top of moir\'e translation and spinless time-reversal symmetry $\hat{\mathcal{T}}$ (TRS), which flips valleys and applies complex conjugation, the system is invariant under three-fold rotations $\hat{C}_{3z}$ as well as $\hat{C}_{2y}$, which flips valley and the top and bottom layers. Note that the action of $\hat{C}_{2z}$ maps $\bm{\delta}\rightarrow-\bm{\delta}$ and is therefore not a symmetry of $\H$. As we have ignored the small `Pauli' twists in the Dirac terms of Eq.~\ref{eq:nonint_BM} and used the leading harmonic and two-center approximation in the tunneling terms \cite{carrMinimalModelLowenergy2019}, there is a particle-hole-inversion symmetry $\mathcal{IC}$ that also exchanges layers \cite{devakul2023magicangle}. In flavor space, $\H$ possesses a global $U(2)_K\times U(2)_{\bar{K}}$ symmetry, which includes charge-$U(1)_c$ and valley-$U(1)_v$ conservation as well as independent $SU(2)_s$ spin-rotations within each valley.

We will primarily be interested in Hartree-Fock, Slater-determinant states because they arise in strong coupling perturbation theory and in Hartree-Fock numerics. Such states are characterized by a projector
\begin{equation}\label{eq:projector}
    P_{\bm{k}\tau s n;\bm{k}'\tau' s' n'}=\langle \hat{c}^\dagger_{\bm{k}\tau s n}\hat{c}^{\phantom{\dagger}}_{\bm{k}'\tau' s' n'}\rangle, \quad P = \frac{1}{2}(1+Q),
\end{equation}
where $n,n'$ are band indices either in the BM band basis or the Chern-sublattice basis (in which case we use $\tilde{\sigma}$ instead of $n$). The filling is determined by $\text{Tr}P=(\nu+4)N_M$, with $N_M$ the number of moir\'e unit cells. We will also find it convenient to use the matrix $Q$, which squares to $1$ since $P$ is a projector.

In the chiral limit~\cite{tarnopolskyOriginMagicAngles2019}, $\kappa = 0$, and at the magic angle, the system has an enhanced symmetry. To see this, we note that at $\kappa=0$, chiral symmetry enforces that the form factors are diagonal in the sublattice basis and the dispersion $h(\bk)$ vanishes at the magic angle. The Hamiltonian is then just the interaction term, which has $U(2) \times U(2) \times U(2) \times U(2)$ symmetry consisting of independent spin and charge rotations within each sublattice and valley degree of freedom. This limit will be the starting point of the strong-coupling theory in Sec.~\ref{sec:strong_coupling}.

\subsection{Hartree-Fock calculations}

For the Hartree-Fock (HF) calculations, we only include fillings $\nu\geq0 $ since the physics at negative fillings is related by particle-hole symmetry given our assumptions on $H^\text{BM}$. The only other conditions we impose on Eq.~\ref{eq:projector} are spin collinearity $s=s'$, and restricted translation symmetry-breaking (TSB). For the latter, the system is allowed to expand its unit cell to double or triple the periods along the moir\'e axes. To ensure convergence in phase diagrams, each parameter involves $>300$ initial seeds of different types, and we use the optimal damping algorithm to accelerate convergence~\cite{cancesODA2000}. For all plots of the HF band structures, the energies are measured with respect to the Fermi level.

\section{Strong-Coupling States at Integer Fillings}\label{sec:strong_coupling}

In this section we report on a strong-coupling approach~\cite{ledwithStrongCouplingTheory2021} to h-HTG, where the non-interacting dispersion and chiral symmetry breaking are taken as perturbations. Detailed calculations are provided in App.~\ref{app:strong_coupling}; here we summarize the structure of the results, the physical intuition behind them, and their implications.

Strong-coupling states, or generalized quantum Hall ferromagnets, Pauli block the density mediated scattering between Bloch states so that the only contribution of the interaction term comes from the overall charging energy associated with reciprocal lattice wavevectors $\bG$. Let us briefly review this argument. Consider states $\ket{\Psi_0}$ that fully fill some combination of sublattice and valley polarized bands, characterized by Hartree-Fock projectors $P = \frac{1}{2}(1+Q)$ that are diagonal in sublattice and valley and $\bk$-independent; here $Q$ is an $8 \times 8$ matrix with eigenvalues $\pm1$ that correspond to filled and empty bands respectively. We do not enforce that the state is diagonal in spin, as the entire sphere of spin directions are degenerate for each sublattice and valley under the enhanced symmetry. Then, the density operator at wavevector $\bq \neq  \bG$ cannot scatter within a flavor due to Pauli blocking, and cannot scatter between flavors because the form factor is diagonal in this basis. We therefore conclude that $\ket{\Psi_0}$ is annihilated by $\rho_{\bq \neq \bG}$. The strong coupling states are eigenvectors of the density operator at reciprocal lattice wavevector $\bG$, and therefore eigenvectors of the interaction Hamiltonian. The associated eigenvalue of the interaction Hamiltonian can be interpreted as a classical ``Hartree'' charging energy that measures how well the charge density of the state $\ket{\Psi_0}$ cancels against the background charge $-4 \overline{\rho}_\bG$. 

The Hartree charging energy yields a splitting of the unperturbed strong coupling states that depends both the filling relative to charge neutrality, $\nu = \frac{1}{2}\tr Q$, and the sublattice polarization $\nu_z = \frac{1}{2} \tr Q \sigma_z$. Explicitly, we have
\begin{equation}
    \frac{E_H}{N_M} = \nu^2\Omega_{00}  + 2\nu \nu_z \Omega_{0z} + \nu_z^2\Omega_{zz}, 
    \label{eq:main_subsplit}
\end{equation}
where
\begin{equation}
\begin{aligned}
    \Omega_{00} & = \frac{1}{2A_M} \sum_\bG V_\bG \ov{\rho}_\bG \ov{\rho}_{-\bG}, \\
    \Omega_{0z} & = \frac{1}{2A_M} \sum_\bG V_\bG \ov{\rho}_\bG \ov{\rho}^z_{-\bG},\\
    \Omega_{zz} & = \frac{1}{2A_M} \sum_\bG V_\bG \ov{\rho}^z_\bG \ov{\rho}^z_{-\bG}.
    \end{aligned}
\end{equation}
Here $\ov{\rho}^z_{\bG} = \sum_\bk \Lambda^{\tau z}_\bG(\bk) = \frac{1}{2}(\rho^{A}_\bG - \rho^B_\bG)$ is the difference in density between sublattices, which does not depend on valley $\tau$ due to time reversal symmetry, and $\Lambda^{\tau z}_\bG(\bk) = \frac{1}{2}(\Lambda^{\tau A}_\bG(\bk) - \Lambda^{\tau B}_\bG(\bk))$. At charge neutrality, states with $\nu = \nu_z = 0$ are exact zero modes of the interaction Hamiltonian, since here the background density is perfectly cancelled. Note that in practice $\Omega_{0z} \approx 0.02\,\text{meV}$ and $\Omega_{zz} \approx 0.003\,\text{meV}$ such that the splitting between states with different $\nu_z$ is small.

The appearance of sublattice polarization in the strong-coupling Hartree energetics is in contrast to TBG, where the charging energy only depends on filling due to an approximate particle-hole symmetry that relates the density of the two sublattices at the same position $\br$. Furthermore, strong-coupling intervalley-coherent (IVC) states are competitive in TBG~\cite{bultinckGroundStateHidden2020,lianTBGIVExact2020}: both valleys of TBG have bands with $C=\pm 1$, and each Chern sector has an approximate $U(4)$ symmetry that rotates not only spin but also valley, thereby relating valley polarized and IVC states~\cite{bultinckGroundStateHidden2020,VafekKangSymmetry,bernevigTBGIIIInteracting2020}. There can be no such symmetry in h-HTG; in fact, all four Chern numbers of the sublattice-valley bands are distinct such that off-diagonal orders must have vortices in the mBZ, equal in number to the difference in Chern number, where the order parameter vanishes~\cite{bultinckMechanismAnomalousHall2020}.

\begin{figure}[t]
    \includegraphics[width=1\columnwidth]{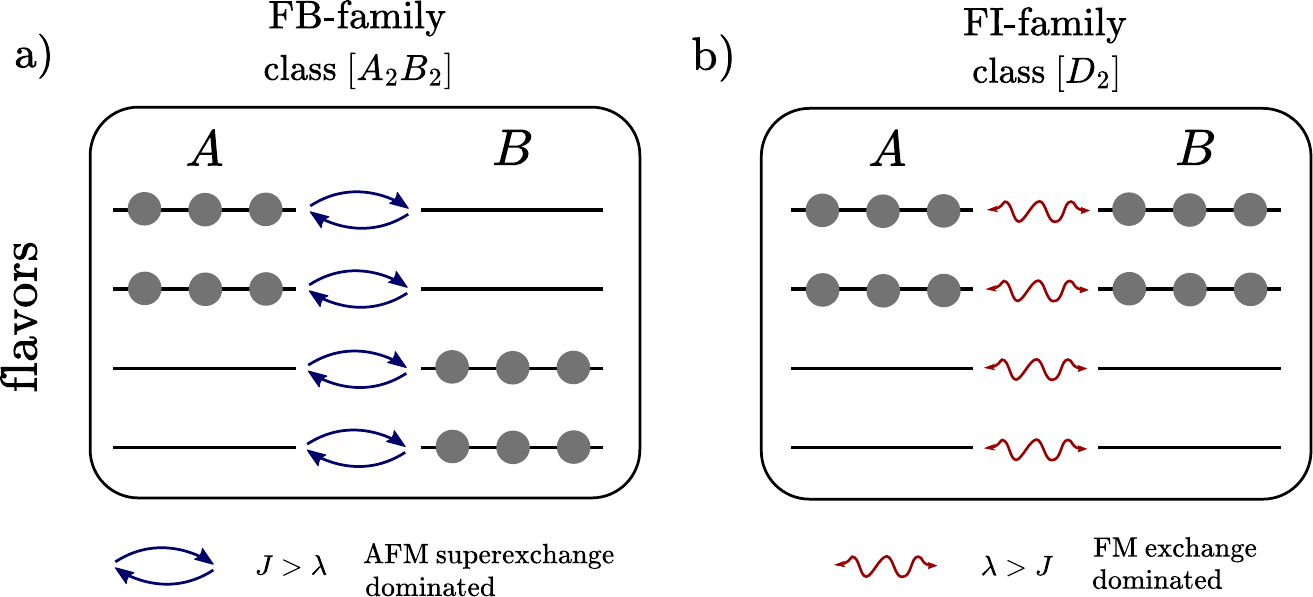}
    \caption{\textbf{Strong-coupling sublattice superexchange and exchange.} a) An example of a strong-coupling state at $\nu=0$ in the flavor-balanced (FB) family that maximally benefits from the dispersion-induced superexchange $J$. b) An example in the flavor-imbalanced (FI) family that maximally benefits from the exchange $\lambda$ that arises due to a finite chiral ratio. Note that we do not explicitly label the spin-valley flavors to reflect the ``flavor permutation symmetry'' for Slater determinants (see the discussion around Eq.~\ref{eq:meanfield_decoupleflavors}).}
    \label{fig:AFM_vs_FM}
\end{figure}

The strong-coupling states are further split upon including both the chiral symmetry breaking, sublattice off-diagonal, part of the form factor, $\Lambda^o_\bq(\bk) = \frac{1}{2} (\Lambda - \tilde{\sigma}_z \Lambda^o_\bq(\bk) \tilde{\sigma}_z )$, which is nonzero for $\kappa > 0$, as well as the dispersion $h(\bk)$. The chiral symmetry breaking form factor leads to exchange energy penalties for states that do not completely fill a spin-valley flavor. Indeed, within a single spin and valley, if only the $A$ sublattice is filled, say, then the form factor can scatter 
$A$ electrons to sublattice $B$ and back, leading to an exchange penalty. This effect, which contributes to first order in perturbation theory, favors the spin and valley flavors to be fully filled or fully empty such that this exchange is Pauli blocked. Another way to characterize this is ``ferromagnetism between sublattices'' induced by exchange: (de-)occupation of the $A$ sublattice within a flavor favors (de-)occupation of the B sublattice within the same flavor. In contrast, the dispersion $h(\bk)$ has a part that tunnels electrons between the two sublattices which vanishes in first order in perturbation theory, but at second order favors ``antiferromagnetism between sublattices'' through ``superexchange'', similar in spirit to the antiferromagnetic exchange between spins induced by hopping in the Hubbard model. In total, we obtain the splitting per moir\'e unit cell
\begin{equation}
    \frac{E_{\text{split}}}{N_M} = \frac{1}{4}(J-\lambda) \tr (Q \sigma_x)^2 = \frac{1}{2}(J-\lambda) \tr Q_A Q_B,
    \label{eq:maintext_splitting}
\end{equation}
where $Q_{A,B}$ are the $4 \times 4$ blocks of $Q$ corresponding to the $A$ and $B$ sublattice respectively, $J \sim h_o^2/U$ is the superexchange scale induced by the off-diagonal dispersion, and $\lambda \sim \abs{\Lambda^o}^2$ is the exchange penalty associated with the off-diagonal form factor. Fig.~\ref{fig:AFM_vs_FM} illustrates these mechanisms for example states at charge neutrality. The derivation of Eq.~\ref{eq:maintext_splitting} is given in App.~\ref{app:strong_coupling}. Note that the (super)exchange scales $J,\lambda\gtrsim1\,\text{meV}$ are significantly larger than the $\nu_z$-splitting scales $\Omega_{0z},\Omega_{zz}$, but smaller than the interaction scale $\approx 30$ meV.

\subsection{Hierarchical labelling of strong-coupling phases}\label{subsec:label_strong}
We outline a compact notation for describing strong-coupling insulators at integer fillings $\nu$, which is indispensable given the multiple symmetries of $\mathcal{H}$. A strong-coupling insulator $\ket{\psi}$ is one that can be obtained via small deformations of a reference state $\ket{\phi}$ consisting of $\nu+4$ fully-occupied Chern bands. By `smooth deformations', we mean that $\ket{\psi}$ and $\ket{\phi}$ share the same symmetries, have comparable Chern band occupations $\langle n_{\tau s \tilde{\sigma}}\rangle$ (where $n_{\tau s \tilde{\sigma}}=\sum_k d^\dagger_{k\tau s\tilde{\sigma}}d_{k\tau\tilde{\sigma}}$), and can be connected without closing the gap. Hence the Chern number is determined simply by summing the corresponding $C_{\tau s \tilde{\sigma}}$ of the filled Chern bands (Eq.~\ref{eq:Chern_numbers}). Based on the arguments of the previous section, we expect that such strong coupling states are the ground state, at least sufficiently close to the chiral limit at the magic angle.

Moving forward we will make the assumption that spin in the $z$ direction is conserved, i.e.~$[Q,S_z]=0$ such that $Q$ is diagonal in spin. While generic strong coupling-states need not satisfy this, we argue in Appendix~\ref{app:spinrot} that all ground states are symmetry related to a state with conserved $S_z$. 
We can then label strong-coupling ground states with the sublattice filling for each $S^z$ spin and valley. The projector is then $P = \diag(N_{\tilde{\sigma}}^{\tau s})$, an $8 \times 8$ diagonal matrix with $N_{\tilde{\sigma}}^{\tau s} = 0,1$ labeling whether the band is filled or empty. We will use $\alpha=(\tau,s)$ as an combined index for both spin and valley, so that $\nu -4 = \sum_{\alpha {\tilde{\sigma}}} N_{{\tilde{\sigma}}}^{\tau s}$, and $\nu_z = \sum_{\alpha {\tilde{\sigma}}} {\tilde{\sigma}} N_{{\tilde{\sigma}}}^{\tau s}$. 

\begin{figure}[t]
    \includegraphics[width=0.8\columnwidth]{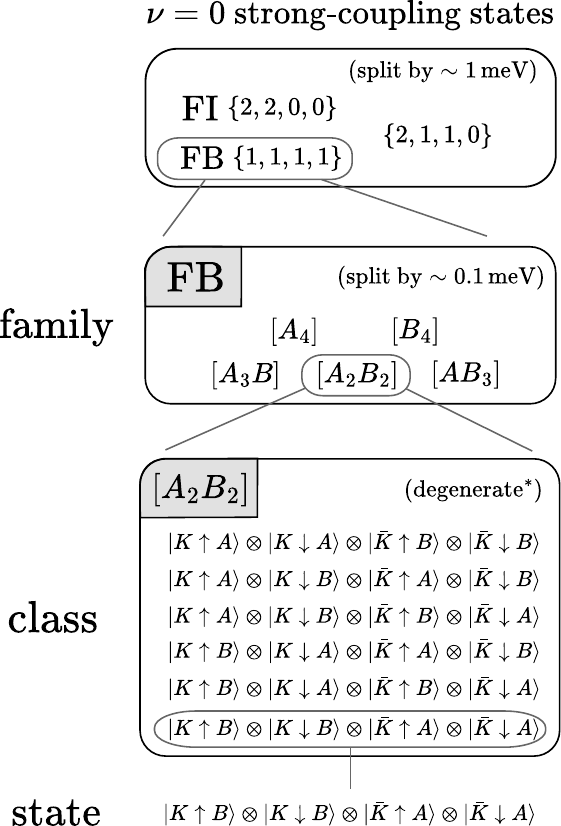}
    \caption{\textbf{Hierarchy of strong-coupling states at $\nu=0$.} The figure illustrates how the particular state $\ket{\psi}=\ket{K\uparrow A}\otimes\ket{K\downarrow A}\otimes\ket{\bar{K}\uparrow B}\otimes\ket{\bar{K}\downarrow B}$ fits into the broader energetic hierarchy of strong-coupling states. 
    Starting from the bottom of the figure, $\ket{\psi}$ is part of the $[A_2B_2]$ class whose states share the same occupations modulo relabelling of spin-valley labels (i.e.~two flavors with an occupied $A$ band and two flavors with an occupied $B$ band), and are degenerate within mean-field theory for our Hamiltonian (Eq.~\ref{eq:intham_main}).  $[A_2B_2]$ in turn is part of the flavor-balanced (FB) family  at $\nu=0$ which has equal occupation $\{1,1,1,1\}$ of all flavors. Different classes are split depending on the sublattice polarization (Eq.~\ref{eq:main_subsplit}). FB and two other families (split by the differing flavor occupations according to Eq.~\ref{eq:maintext_splitting}) together comprise the manifold of $\nu=0$ strong-coupling states.}
    \label{fig:strong_coupling_schematic}
\end{figure}

The separation of energy scales indicated by the strong-coupling analysis suggests a natural hierarchical labelling scheme of strong coupling states, summarized by Fig.~\ref{fig:strong_coupling_schematic}.  

The most significant splitting is due to the sublattice exchange-superexchange term (Eq.~\ref{eq:maintext_splitting}), which separates states into ``families'' separated by energies of order $\sim 1$meV.
For $J<\lambda$, the 
term $\tr Q_A Q_B = \sum_\alpha (N_A^{\alpha} - \frac{1}{2})(N_B^{\alpha} - \frac{1}{2})$ in Eq.~\ref{eq:maintext_splitting} favors as many double and empty occupations, $N_A^\alpha = N_B^\alpha$, as possible, while
for $J>\lambda$, it is best for flavors to be as singly occupied as possible (see Fig.~\ref{fig:AFM_vs_FM}). 
This term dictates the overall distribution among the flavors, agnostic to the details of the specific spin/valley/sublattice bands being filled.
We therefore label each family by the flavor occupation numbers $\{N_{f_0},N_{f_1},N_{f_2},N_{f_3}\}$, where $N_{f_i}\in 0,1,2$ are listed in descending order.
Different families are distinguished based on the number of fully-filled, singly-occupied, and empty flavors.

As discussed above, the family that doubly occupies as many flavors as possible is favored for $\lambda > J$; we will call this family ``maximally flavor imbalanced'' (FI). Meanwhile, the family that has as few doubly occupied flavors as possible is called ``maximally flavor balanced'' (FB), and is favored for $J>\lambda$. 
For $\nu=0$, the FI and FB family correspond to the $\{2,2,0,0\}$ and $\{1,1,1,1\}$ labels (illustrated in Fig.~\ref{fig:AFM_vs_FM}). Note that $\nu=3$ only has one family, $\{2,2,2,1\}$, which we arbitrarily designate as FB. 
The description in terms of flavor occupations is also useful beyond the strong-coupling regime, and can be applied to TSB phases and non-integer fillings.

The states within a given family are further separated into distinct ``classes'' based on their overall sublattice polarization $\nu_z$ (Eq.~\ref{eq:main_subsplit}), which are separated by smaller energies of order $\sim 0.1$meV.
A descriptive labelling of a particular class is achieved by the notation $[D_{N_D} A_{N_A} B_{N_B}]$, where $N_A$ ($N_B$) counts the flavors where only the $A$ ($B$) band is filled, and $N_D$ counts the fully-filled flavors where both $A$ and $B$ are occupied.
These integers satisfy $N_A+N_B+2N_D=\nu+4$.
We consider some examples:
For $\nu=3$, the class $[D_3 A_0 B_1]$ (or, $[D_3 B]$ for short), refers to any state where all bands are filled except for one of the $A$ bands: $(K\uparrow A)$,$(K\downarrow A)$,$(\bar{K}\uparrow A)$, or $(\bar{K}\downarrow A)$.
At charge neutrality $\nu=0$, the class $[A_2 B_2]$ means two flavors have only the $A$ band occupied, and the other two flavors have only the $B$ band occupied.  
Note that the valley degree of freedom allows for multiple possible Chern numbers --- the class $[A_2 B_2]$ includes states with $C=-6,0,6$.

The states within a class are exactly degenerate within our strong-coupling analysis and self-consistent HF theory.
This can be understood due to the presence of a ``flavor permutation symmetry'' that exists for Slater determinant states of h-HTG that are $S^z$ and valley conserving.
For instance, the strong-coupling energetics outlined in the previous paragraph are invariant under permutations of the $\alpha = (\tau, s)$ index.
This is a consequence of the fact that the Fock energy decomposes into a sum over contributions from each flavor,
\begin{equation}
E_F[P] = \sum_{\alpha} E_{F\alpha}[P_\alpha],
\label{eq:meanfield_decoupleflavors}
\end{equation}
where $E_{F\alpha}$ for different $\alpha$ are symmetry related,
while the Hartree energy only depends only on the total density (see App.~\ref{app:flavor_permutation}).
This flavor permutation symmetry can result in very physically distinct states having equivalent HF energies.
For example, consider a quantum spin Hall (QSH) insulator at $\nu=+2$, consisting of filling all Chern-sublattice bands except for $(K\uparrow A)$ and $(\bar{K} \downarrow A)$. By exchanging the flavors $(\bar{K},\downarrow) \leftrightarrow (K, \downarrow)$, through TRS applied on spin $\downarrow$, we arrive at a valley polarized $|C| = 2$ Chern insulator (CI).

The application of time reversal to only one spin species only makes sense at mean-field level as a result of the decoupling Eq.~\ref{eq:meanfield_decoupleflavors}; this ``Hartree-Fock symmetry'' does not make sense either as a unitary or antiunitary symmetry on the full Hilbert space, and the degeneracy between the CI and QSH is expected to be lifted by quantum fluctuations outside of mean field theory. From the strong coupling perspective, they could be split in sufficiently high order perturbation theory in the dispersion $h(\bk) \neq 0$; in the leading order calculation of $J$ described in the previous subsection, the perturbed state is still a Slater determinant (see App.~\ref{app:strong_coupling}). We note that while some of the flavor permutation induced degeneracies can be lifted by including terms like intervalley-Hunds, which reduces $SU(2)_K \times SU(2)_{\bar{K}} \to SU(2)_s$, the degeneracy between the QSH and CI only depends on time reversal and requires going beyond mean-field-theory. There is another mechanism that is expected to favor the QSH when considering the three-dimensional nature of real space, which is that the CI generally has a finite orbital magnetization and hence a magnetic field energy cost.
Although these magnetic fields vanish in an infinite 2D system, recall that h-HTG appears only as a finite domain of the full HTG super-moir\'e structure, and so such effects may be non-negligible.
We leave a detailed analysis of such effects for future work.

To illustrate the multitude of possible states due to flavor permutations alone, we consider concrete examples for one class in each filling in the FB family. For the $[A_2B_2]$ class at $\nu=0$, filling an $A$ and a $B$ band in each valley leads to $C=0$. However, these states still have valley Chern number $|C_v|=|C_K-C_{\bar{K}}|=2$, and hence realize quantized valley Hall (QVH) insulators. If $S_z$ is a good quantum number, then the spin Chern number can take values $|C_s|=0,3$, potentially allowing for a quantized spin Hall (QSH) effect. On the other hand, a valley-sublattice-locked state where the occupied $A$ ($B$) bands are in valley $K$ ($\bar{K}$) would lead to $C=6,C_v=-2$ and $C_s=0$.
For $\nu=1$, the possible sets of topological indices of the $[DB_3]$ class are more constrained. For instance, doubly occupying the $K\uparrow$ flavor results in $C=1,C_v=-7,C_s=1$---other possibilities lead to identical magnitudes of the various Chern numbers. The states of the $[D_2B_2]$ class at $\nu=2$ are primarily distinguished by whether the doubly occupied bands share the same valley. For such valley polarized states, we have $|C|=2,|C_v|=4$, and $C_s=0$, while the valley-unpolarized representatives have $C=0$ and $|C_v|=6$. Finally, the $[D_3B]$ class at $\nu=3$ always has $|C|=|C_v|=|C_s|=1$. Note that the odd integer states in the FB family necessarily have $|C|>0$ and a net valley and spin polarization.

\begin{figure*}[t]
    \includegraphics[width=\textwidth]{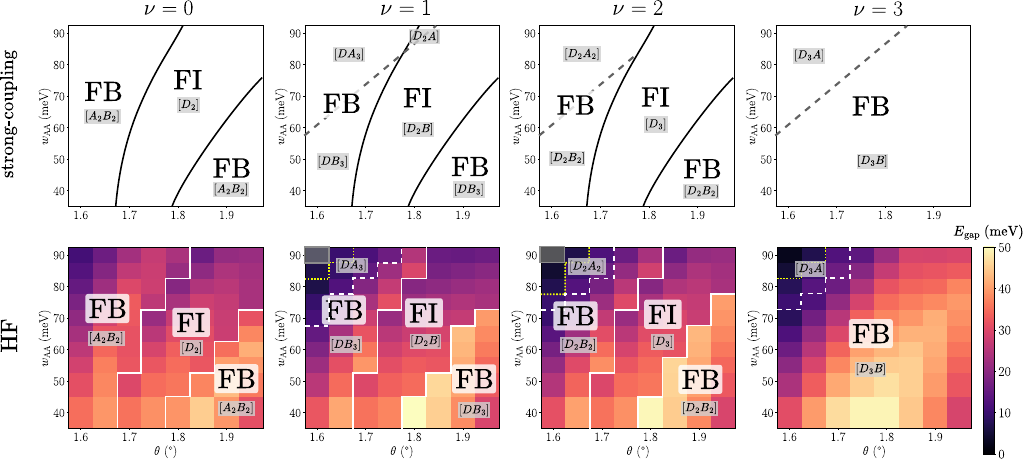}
    \caption{\textbf{Phase diagrams for non-negative integer fillings $\nu$ at zero displacement field.} $w_{\text{AA}}$ denotes same-sublattice interlayer tunneling ($w_{\text{AB}}$ is fixed at $110\,\text{meV}$). $\text{FB}$ and $\text{FI}$, which stand for maximally flavor-balanced and maximally flavor-imbalanced respectively, represent different families of nearly-degenerate strong-coupling phases that are distinguished by the patterns of flavor occupations (see Tab.~\ref{tab:strong_coupling}). Bracketed labels $[D_{N_D}A_{N_A}B_{N_B}]$ with grey background denote the precise strong-coupling class with minimal energy. $N_D,N_A,N_B$ refer to the number of flavors that are fully filled, have only the $A$ band filled, or have only the $B$ band filled, respectively.  
    \textbf{Top:} Phase diagrams derived from strong-coupling analysis.  Black solid lines separate regions that favor the FB (FI) family due to $J>\lambda$ ($J<\lambda$), see Eq.~\ref{eq:maintext_splitting}. For $\nu>0$, grey dashed lines separate regions that favor polarization into the $A$ ($B$) sublattice due to $\Omega_{0z}<0$ ($\Omega_{0z}>0$), see Eq.~\ref{eq:main_subsplit}.
    \textbf{Bottom:} Numerical self-consistent Hartree-Fock (HF) phase diagrams. Color plot shows the HF band gap $E_\text{gap}$. White solid lines indicate dominant boundaries, shaded grey areas denote absence of a charge gap, and dotted yellow lines indicate a weak-coupling excitonic instability. Dashed white lines indicate transitions between strong-coupling classes within the same family. System size is $12\times12$, and relative permittivity $\epsilon_r=8$.}
    \label{fig:wAA_theta_phase_diagram}
\end{figure*}

\subsection{Strong-coupling phase diagram}

In the top row of Fig.~\ref{fig:wAA_theta_phase_diagram}, we show the strong-coupling prediction for the phase diagrams as a function of $w_\text{AA}$ and $\theta$. The main phase boundaries (solid lines) reflect the competition between the different families FB and FI, which is controlled by the relative values of $J$ and $\lambda$ (Eq.~\ref{eq:maintext_splitting}). 
Within each family, there are also secondary phase boundaries (dashed lines) separating distinct classes, based on the sublattice polarization $\nu_z$ if some flavors are partially occupied (Eq.~\ref{eq:main_subsplit}). 
At charge neutrality $\nu=0$, we always have $\nu_z=0$ since $\Omega_{zz}$ is positive. 
At finite integer fillings, the favored sublattice is set by the sign of $\Omega_{0z}$ (since, for our range of parameters, $\Omega_{zz}$ is usually very small); while $\Omega_{0z}$ is positive (favoring the $B$ sublattice) for much of the phase diagram considered, we find it becomes negative at small twist angles and large $w_{AA}$.
These findings are in excellent agreement with the self-consistent HF calculations discussed in the next section.

\section{Hartree-Fock phase diagram at integer fillings}\label{sec:phase_wAA}

The bottom row of Fig.~\ref{fig:wAA_theta_phase_diagram} shows the integer HF phase diagrams as a function of $w_{AA}$ and twist angle $\theta$. Almost all regions show a non-zero HF gap $E_\text{gap}$, indicating the presence of correlated insulators for the chosen parameters. As expected from the narrow BM dispersion and strong interactions, the insulators are all strong-coupling phases (see Sec.~\ref{subsec:label_strong}), as confirmed by the substantial polarization in flavor and Chern-sublattice space. The positions of the phase boundaries (white solid and dashed lines) are remarkably similar across all fillings. For $\nu=0,1,2$, the phase diagrams are dominated by two different strong-coupling classes, whose sublattice-flavor occupations are indicated with a grey background. 
One of them is the ground state for a window of twist angles near the magic angle $\theta=1.8^\circ$, while the other emerges for slight detuning away from this. For $\nu=1,2,3$, another strong-coupling class appears in the top-left corner.

\begin{figure}[t]
    \includegraphics[width=0.9\columnwidth]{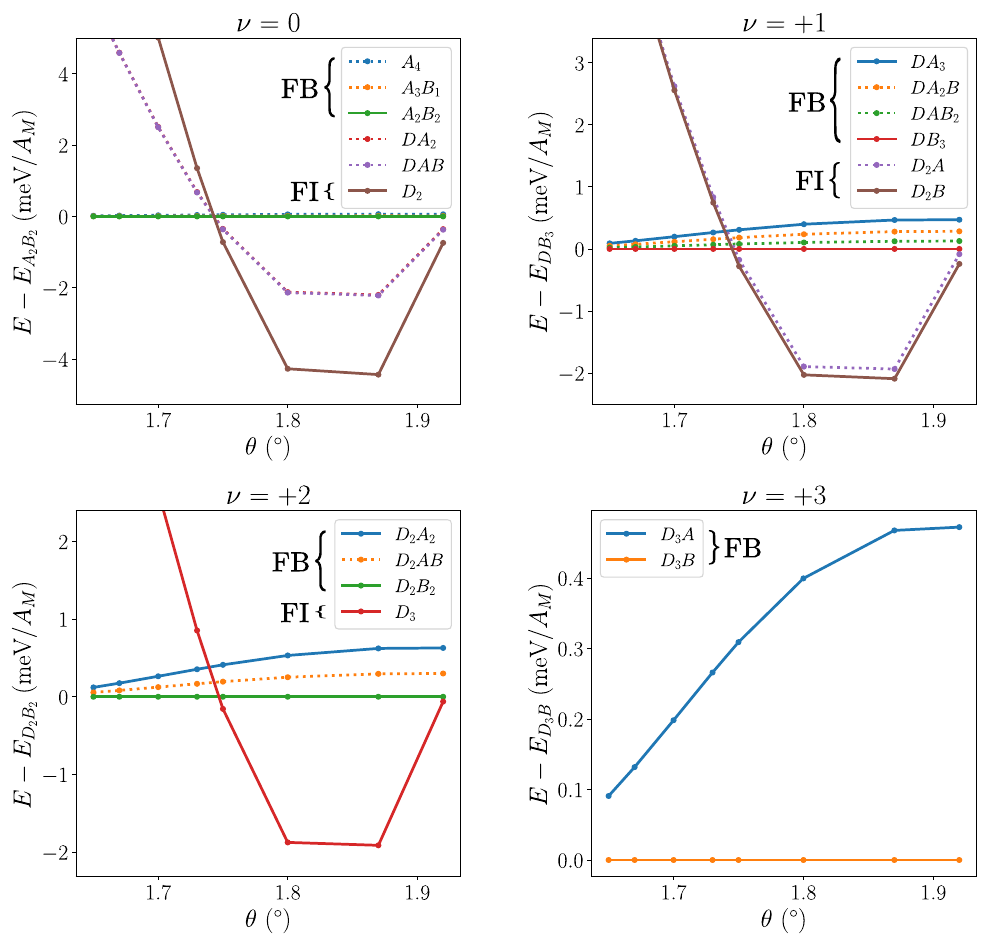}
    \caption{\textbf{Energetic competition between strong-coupling phases at zero displacement field.} Self-consistent HF energies of different strong-coupling classes. They are labelled according to the configuration of occupied sublattice bands, e.g.~$DB_3$ indicates one fully-filled flavor and three additional filled $B$ bands in singly-occupied flavors. Phases that appear in Fig.~\ref{fig:wAA_theta_phase_diagram} are denoted with solid lines. States with the same number of fully occupied flavors belong to the same family (see Tab.~\ref{tab:strong_coupling}) and are nearly degenerate. For $\nu=0$, we have omitted classes which are related by particle-hole symmetry to the ones shown (e.g.~$E_{DA_2}=E_{DB_2}$). Note that no classes in the figures are degenerate with each other, though the splittings at $\nu=0$ are almost invisible on this scale. System size is $12\times 12$, $w_{\text{AA}}=70\,\text{meV}$.}
    \label{fig:strong_coupling_energies}
\end{figure}

\begin{table}
\centering
\newcommand{\colskip}{\hskip 0.15in}
\renewcommand{\arraystretch}{1.15}
\begin{tabular}{
l @{\hskip 0.2in} 
l @{\colskip} 
l @{\colskip} 
l  }\toprule[1.3pt]\addlinespace[0.3em]
Family & 
$\nu$ & 
\{flavor occ.\} & 
$|C|$
\\ \midrule
FB & 0 & $\{1,1,1,1\}$ & $\bm{0},\bm{6},\mathsmaller{3}$ \\
 & 1 &  $\{2,1,1,1\}$ & $\bm{1},\mathsmaller{2},\mathsmaller{4},\mathsmaller{5}$ \\
 & 2 & $\{2,2,1,1\}$ & $\bm{0},\bm{2},\mathsmaller{1},\mathsmaller{3},\mathsmaller{4}$ \\
 & 3 & $\{2,2,2,1\}$ & $\bm{1},\mathsmaller{2}$ \\
\midrule
FI & 0 & $\{2,2,0,0\}$ & $\bm{0},\bm{2}$ \\
 & 1 &  $\{2,2,1,0\}$ & $\bm{0},\bm{2},\mathsmaller{1},\mathsmaller{3}$ \\
 & 2 &  $\{2,2,2,0\}$ & $\bm{1}$ \\
\midrule
 & 0  & $\{2,1,1,0\}$ & $\mathsmaller{0},\mathsmaller{1},\mathsmaller{2},\mathsmaller{4}$ 
\\\bottomrule[1.3pt]
\end{tabular}
\caption{\textbf{Families of strong-coupling states.} At a fixed filling, each family includes different classes distinguished by the sublattice occupations (see main text). $\{\text{flavor occ.}\}$ lists the flavor occupations in descending order. The FB (FI) family minimizes (maximizes) the number of fully-filled flavors. $C$ denotes the possible Chern numbers, with large bold entries corresponding to the dominant ground-state HF phases obtained in Fig.~\ref{fig:wAA_theta_phase_diagram}.\label{tab:strong_coupling}}
\end{table}
 
For each parameter, our HF calculations produce a particular strong-coupling class $[D_{N_D}A_{N_A}B_{N_B}]$ with the lowest energy. 
However, Fig.~\ref{fig:strong_coupling_energies} shows that the energies of multiple strong-coupling classes can be closely competitive.
Recall that even within a class, there are multiple distinct patterns of symmetry-breaking and Chern numbers. At each filling, the classes group into families depending on the number of fully occupied flavors $N_D$, which determines the energetics under exchange $\lambda$ and superexchange $J$, as described by Eq.~\ref{eq:maintext_splitting}. 
The family with the maximum possible $N_D=\lfloor \frac{\nu +4}{2}\rfloor $ is denoted FI (``maximally flavor imbalanced''), and is favored near the magic angle where the non-interacting bandwidth is smallest such that exchange outweighs superexchange $\lambda > J$. In contrast, the family with the minimum possible $N_D$, denoted FB (``maximally flavor balanced''), is favored for larger bandwidths where superexchange between sublattices outweighs exchange. For $\nu=0$, there is also an intermediate family with $N_D=1$, which we do not name since it does not appear as the ground state in Fig.~\ref{fig:wAA_theta_phase_diagram}. The possibilities are summarized in Tab.~\ref{tab:strong_coupling}. These families are separated by energies $\gtrsim 1\,\text{meV}$ except near the phase boundaries. The dependence of the FB vs.~FI competition on chiral ratio and twist angle matches closely with the perturbative strong-coupling analysis. However, consistent with the strong-coupling analysis of Eq.~\ref{eq:main_subsplit}, the splittings within each family from distinct $\nu_z$ are significantly smaller, especially at $\nu=0$ where the differences are $\lesssim 0.05\,\text{meV}$. These sensitive near-degeneracies can easily be affected by details of the modeling, as well as extrinsic effects such as sublattice coupling to the hBN substrate. Therefore, while our prediction of the lowest energy strong-coupling family is robust and in excellent agreement with strong-coupling perturbation theory in Sec.~\ref{sec:strong_coupling}, the particular strong-coupling class and state that ultimately emerges may be sensitively detail-dependent.
To reflect this, the phase diagrams of Fig.~\ref{fig:wAA_theta_phase_diagram} also label the relevant strong-coupling family. At $\nu=+3$, there is only a single option for the family since we must have $N_D=3$. In Tab.~\ref{tab:strong_coupling}, we also list the possible Chern numbers, with the bold entries corresponding to the lowest-energy representatives over major parts of the phase diagrams.

The emergence of symmetry-broken Chern insulators should give rise to several characteristic features in experimental observables. 
In devices consisting of a single h-HTG region spanning the electrical contacts, these states spontaneously break TRS and will exhibit a quantized anomalous Hall response.
More likely, however, experimental detection of such signatures in transport will be complicated by the super-moir\'e structure of HTG which consists of h-HTG domains and their $\hat{C}_{2z}$-related $\bar{\text{h}}$-HTG counterparts~\cite{devakul2023magicangle}; we defer a detailed discussion to Sec~\ref{sec:discussion}
The Chern insulators can also be uncovered by applying a perpendicular magnetic field and studying the $\nu-B$ plane, where such states appear as sloped lines according to the Streda formula. This method allows for the identification of multiple competing states which are rooted at the same integer filling but have different Chern numbers $C$ (Tab.~\ref{tab:strong_coupling}), and is also accessible to probes such as STM~\cite{wongCascadeElectronicTransitions2020,choi2021correlation,choi2021interactiondriven,nuckolls2023quantum,kim2023imaging,turkel2022orderly,nuckollsStronglyCorrelatedChern2020,zhang2023local} and SET~\cite{xieFractionalChernInsulators2021,yu2022correlated,yu2022spin} which can map out the local moir\'e-scale physics.

\begin{figure}[t]
    \includegraphics[width=1\columnwidth]{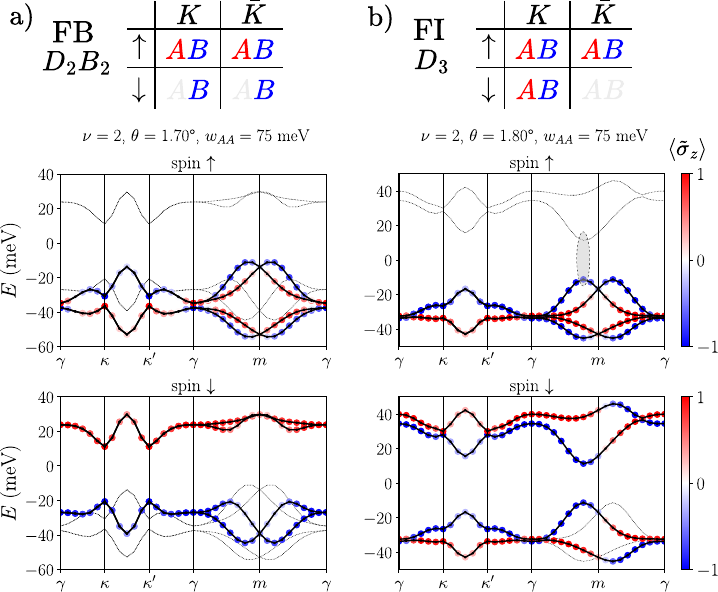}
    \caption{\textbf{Strong-coupling phases at $\nu=+2$.} a) Top shows the occupations in the sublattice-polarized basis for a low-Chern $C=0$ representative of $D_2B_2$ phase (FB family).  Bottom shows the HF band structures in each spin sector, with colored dots indicating the Chern basis polarization. The opposite spin bands are also shown with thin dotted lines. b) Same as a) but for a low-Chern $C=0$ representative of the $D_3$ phase (FI family). Shaded oval indicates an incipient intervalley excitonic instability. System size is $18\times18$.}
    \label{fig:strong_coupling_nu2}
\end{figure}

In many magic-angle graphene systems, the real-space charge inhomogeneity of a filled central band within the moir\'e cell leads to substantial interaction-induced renormalization, especially at finite fillings. For instance, the flat-band wavefunctions of TBG are concentrated at $AA$-stacking regions, and the Hartree-renormalized dispersion experiences a pronounced dip at $\gamma$ for positive fillings, substantially increasing the bandwidth from its non-interacting value~\cite{Guineaelectrostatic2018,ardemakercharge2019,ceapinning2019,goodwin2020hartree,Kangcascades2021,pierce2021unconventional,parkerFieldtunedZerofieldFractional2021}. This complicates the identification of the correct starting point for theoretical treatments of various phenomena. In h-HTG, the presence of two shifted moir\'e patterns (Fig.~\ref{fig:D_nu_splash}) smoothens the charge modulation, which should reduce the impact of such renormalization effects. This expectation is borne out in Fig.~\ref{fig:strong_coupling_nu2}, which plots the self-consistent HF band structure for the $[D_2B_2]$ and $[D_3]$ phases at $\nu=+2$. The bands are color coded by the Chern basis polarization, which reveals the strong-coupling nature of the states. By comparing with the dispersion and sublattice polarization of the non-interacting bands (Fig.~\ref{fig:BM_zeroU}), it is clear that the bands are not significantly deformed, in contrast to other moir\'e systems. For instance, the pronounced dispersion of the $B$ bands along the $\gamma-m$ lines is preserved. 

This `band rigidity' is also conducive towards the stabilization of insulators at non-zero fillings. In TBG, the Hartree corrections progressively degrade the mean-field exchange gap at large fillings, such that the strong-coupling candidates at $|\nu|=3$ have a vanishing/small gap which is sensitive to details of the modelling, and may give way to other candidate states~\cite{kangNonAbelianDiracNode2020,lianTBGIVExact2020,kwan2021kekule,songheavy2022,xienu3_2023}. On the other hand, the HF gap in h-HTG remains similar across all integer fillings (Fig.~\ref{fig:wAA_theta_phase_diagram}), suggesting that the insulating character will be more robust against quantum fluctuations and other deleterious effects like disorder. Note that the FB family has a larger insulating gap than the FI family for $|\nu|>0$, since all of the dispersive $B$ bands are either above or below the Fermi level.

\begin{figure}[t]
    \includegraphics[width=1\columnwidth]{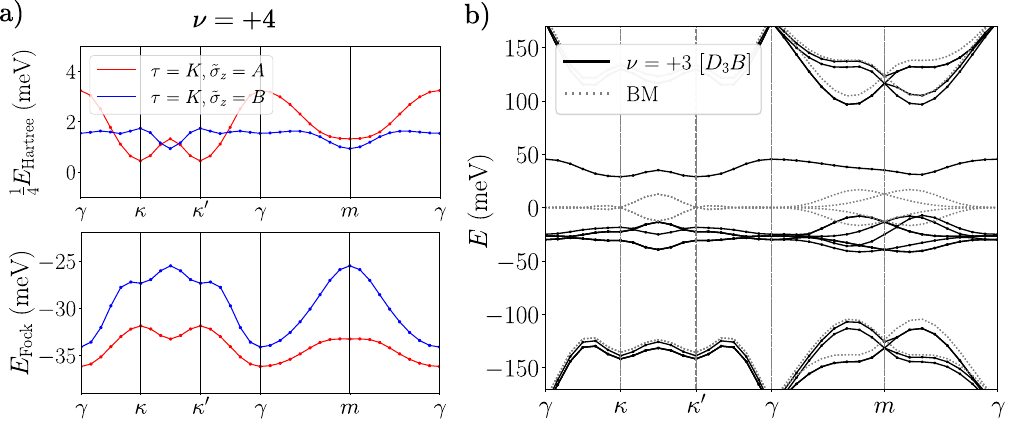}
    \caption{\textbf{Hartree-Fock potentials and remote bands.} a) Hartree (divided by 4) and Fock potentials corresponding to the fully filled state at $\nu=+4$. We show the diagonal component in the Chern-sublattice basis and for valley $K$.  b) Black lines show HF band structure for a state in the $\nu=+3$ $[D_3 B]$ strong-coupling class (see Fig.~\ref{fig:strong_coupling_nu3}a for schematic) when including the lowest remote bands self-consistently. Energies are not measured relative to the chemical potential. Dotted grey lines show the non-interacting BM bands. System size is $18\times18$, $\theta=1.80^\circ$, $w_\text{AA}=75\,\text{meV}$, and $\epsilon_r=8$ for interacting calculations.}
    \label{fig:Fock_potentials}
\end{figure}

To see the interaction renormalization more explicitly, we plot the Hartree and Fock components of the mean-field Hamiltonian corresponding to the fully filled $\nu=+4$ symmetry-preserving state in Fig.~\ref{fig:Fock_potentials}a. The potentials are shown for the diagonal entries in the Chern-sublattice basis, and we normalize $E_\text{Hartree}$ by $1/4$ to estimate the contribution from a single filled band. As in TBG, the direct and exchange terms tend to cancel each other somewhat. However, the Hartree part is significantly suppressed compared to TBG. Fig.~\ref{fig:Fock_potentials}b illustrates the HF band structure for a strong-coupling insulator at $\nu=+3$ where the nearest remote bands have also been included self-consistently. Note that the energy axis has not been zeroed to the chemical potential, and we have subtracted off a classical charging energy arising from $V(0)$. Even though the system is far from neutrality and develops an appreciable exchange splitting, there remains a sizable gap to the remote bands, whose position and shape are qualitatively unchanged. There is negligible remote band mixing, as the central bands retain $\gtrsim99.5\%$ fidelity. This is enabled by the large initial remote gap in the non-interacting dispersion, and the absence of strong momentum-dependent Hartree corrections. Therefore, unlike in many other strongly-interacting moir\'e systems, we anticipate the approximation of restricting to the central bands for interacting calculations to remain quantitatively correct all the way to $|\nu|=4$. The suppression of Hartree should also lead to a smaller overall positive offset to the filling-dependent inverse electronic compressibility $d\mu/dn$.

\begin{figure}[t]
    \includegraphics[width=1\columnwidth]{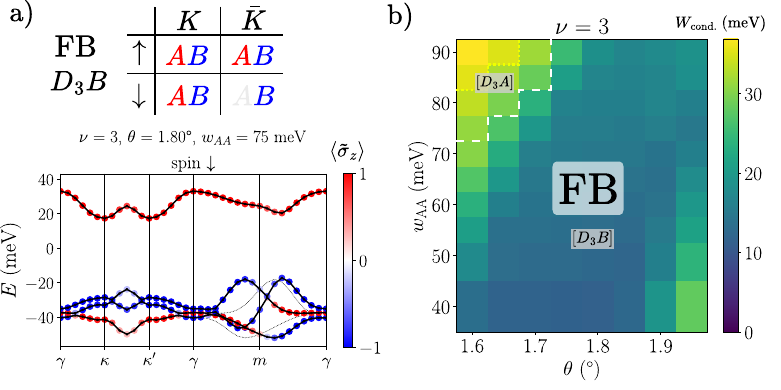}
    \caption{\textbf{Strong-coupling Chern insulator at $\nu=+3$.} a) Top shows the occupations in the sublattice-polarized basis for the $C=1$ state in the $D_3B$ class. Bottom shows the HF band structures in the partially occupied spin sector, with colored dots indicating the Chern-sublattice polarization. The opposite spin bands are also shown with thin dotted lines. System size is $18\times18$. b) HF conduction bandwidth $W_{\text{cond.}}$ as a function of $w_{\text{AA}}$ and $\theta$. System parameters identical to those of Fig.~\ref{fig:wAA_theta_phase_diagram}.}
    \label{fig:strong_coupling_nu3}
\end{figure}

It has been proposed that the combination of ideal quantum geometry and suppressed interaction renormalization in h-HTG is potentially conducive towards realizing fractional Chern insulators, especially at fractional fillings beyond $|\nu|=3$~\cite{devakul2023magicangle}. Fig.~\ref{fig:strong_coupling_nu3}a shows that for most of the phase diagram at $\nu=+3$, the lowest-energy strong-coupling phase is $[D_3B]$, which has a single narrow conduction band in the less dispersive $A$ sublattice with $|C|=1$. In Fig.~\ref{fig:strong_coupling_nu3}b, we chart the conduction bandwidth $W_\text{cond.}$, demonstrating that it remains small $W_\text{cond.}\lesssim 15\,\text{meV}$ over most of the phase diagram. Furthermore, the fact that the conduction band is nearly wholly composed of one sublattice band suggests that it retains its favorable quantum geometry. 

We caution that the energetically-preferred flavor and sublattice polarization at integer fillings may not necessarily reflect the situation at finite doping. For instance, while the mean-field ground state at $\nu=3$ is $[D_3B]$ (we use the terms state and class interchangably here), Fig.~\ref{fig:strong_coupling_energies} shows that the $[D_3A]$ state differs in energy by less than 1~meV per moir\'e cell. Simple considerations of the contrasting dispersion of $A$ and $B$ bands suggest that beyond some finite electron doping the system may experience a first-order transition where the Chern-sublattice polarization switches sign. This is because electron-doping a $[D_3A]$ state involves adding carriers to the unoccupied $B$ band, which contains more significant energy troughs compared to the $A$ band (Fig.~\ref{fig:BM_zeroU}a). Hence doped electrons are less costly if the parent insulator is $[D_3A]$ rather than $[D_3B]$. It is possible then that at some critical filling $3+\delta_c$, this discrepancy is enough to overcome the initial energy difference of the parent insulators. The precise value of $\delta_c$, if this mechanism does indeed occur, is sensitive to details such as the initial energy splitting between $[D_3A]$ and $[D_3B]$ and the correlation energy of the partially filled band. Similar considerations apply between other integer fillings, and may factor into potential `reset' and cascade physics~\cite{zondinerCascadePhaseTransitions2020,wongCascadeElectronicTransitions2020}. In TBG, such effects are often explained with flavor transitions, but here the additional possibility of first-order sublattice transitions complicates the picture.

\begin{figure}[t]
    \includegraphics[width=1\columnwidth]{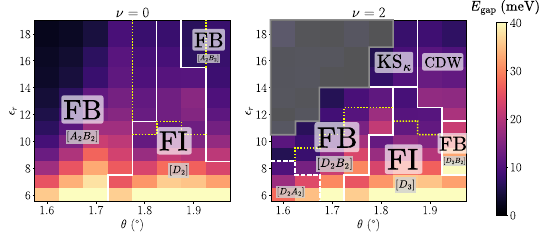}
    \caption{\textbf{Effect of tuning interaction strength at zero displacement field.} Color plot shows the Hartree-Fock band gap $E_\text{gap}$. Labelling is the same as Fig.~\ref{fig:wAA_theta_phase_diagram}. System size is $12\times12$ and $w_\text{AA}=75\,\text{meV}$. [CDW: charge density wave, KS$_\kappa$: Kekul\'e spiral]}
    \label{fig:epsr_theta_phase_diagram}
\end{figure}

Finally, we discuss the impact of tuning the interaction strength. In Fig.~\ref{fig:epsr_theta_phase_diagram}, we plot the phase diagram at even integer fillings as a function of $\epsilon_r$ and $\theta$. Consistent with the FI family favoring a  narrower bandwidth (since dispersion increases $J$), the twist angle window where FI has the lowest energy shrinks in favor of the FB family when weakening interactions (increasing $\epsilon_r$). We also note that when the band gap is sufficiently small, the strong-coupling phases can be susceptible to a weak-coupling excitonic instability, indicated by yellow dotted lines in Figs.~\ref{fig:wAA_theta_phase_diagram} and~\ref{fig:epsr_theta_phase_diagram}. The relevant exciton is composed of electrons and holes at the $B$ band extrema---an example is highlighted in Fig.~\ref{fig:strong_coupling_nu2}b. Depending on the flavor nature of the exciton, this can occur in the intervalley channel and break $U(1)_v$, possibly with a finite (incommensurate) moir\'e wavevector. However, the change in Chern basis occupations is minor, and the resulting state retains most properties of the non-excitonic parent phase. Hence, such effects will be difficult to detect experimentally. For sufficiently weak interactions at non-zero $\nu$, the strong-coupling phases can be replaced by gapped TSB phases such as a commensurate Kekul\'e spiral (KS$_\kappa$) or charge density wave (CDW), which will be elaborated on later in the context of finite displacement fields (Sec.~\ref{sec:finite_displacement}). Sizable regions of the phase diagram can also become gapless, especially for smaller twist angles where the interaction is relatively weaker for fixed $\epsilon_r$ due to the increased moir\'e length.

\section{Finite displacement field}\label{sec:finite_displacement}

\begin{figure}[t]
    \includegraphics[width=1\columnwidth]{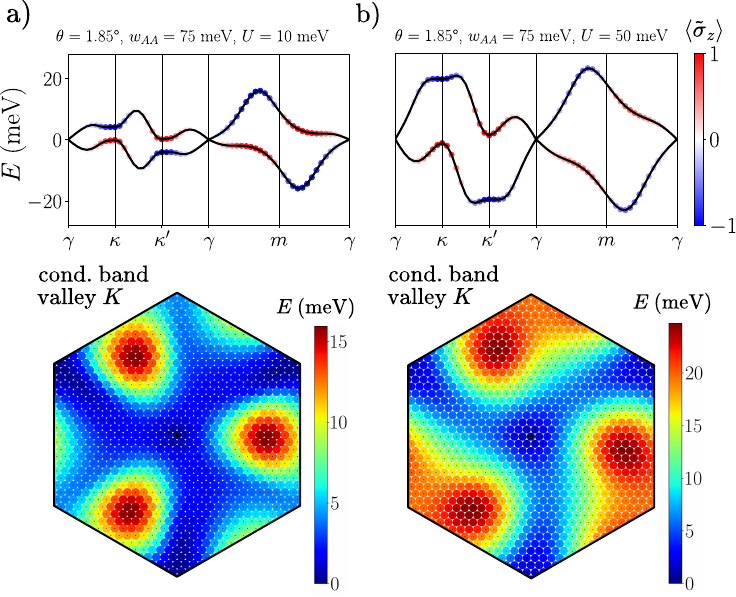}
    \caption{\textbf{Non-interacting band structure at finite displacement field.} a) Top shows dispersion of the central bands with interlayer potential $U=10\,\text{meV}$. Only valley $K$ is shown. Color indicates Chern-sublattice polarization $\langle\tilde{\sigma}_z\rangle$.  Bottom shows energy dispersion of the conduction band in the mBZ. b) Same as a) except with $U=50\,\text{meV}$.}
    \label{fig:BM_finiteU}
\end{figure}

\begin{figure*}[t]
    \includegraphics[width=\textwidth]{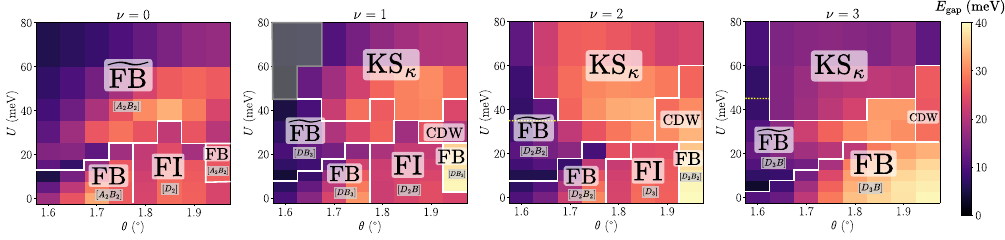}
    \caption{\textbf{Phase diagrams for non-negative integer fillings $\nu$ at finite displacement field.} Color plot shows the Hartree-Fock band gap $E_\text{gap}$. White lines indicate approximate phase boundaries, shaded grey areas denote absence of a charge gap, and dotted yellow lines indicate a weak-coupling excitonic instability. For the translation-invariant phases, we also describe the dominant flavor-sublattice occupations of the lowest-energy HF solution, as in Fig.~\ref{fig:wAA_theta_phase_diagram}. Properties of the phases at finite displacement field are listed in Tab.~[\ref{tab:finite_U_phases}]. System size is $12\times12$, $\epsilon_r=8$, and $w_\text{AA}=75\,\text{meV}$. [CDW: charge density wave, KS$_\kappa$: Kekul\'e spiral]}
    \label{fig:U_theta_phase_diagram}
\end{figure*}

\begin{table}
\centering
\newcommand{\colskip}{\hskip 0.15in}
\renewcommand{\arraystretch}{1.15}
\begin{tabular}{
l @{\hskip 0.2in} 
l @{\colskip} 
l @{\colskip} 
l  }\toprule[1.3pt]\addlinespace[0.3em]
Phase/Family & 
$\nu$ & 
\{flavor occ.\} & 
$|C|$
\\ \midrule
$\widetilde{\text{FB}}$ & 0 & $\{1,1,1,1\}$ & $\bm{0},\bm{2},\mathsmaller{1}$\\
 & 1 & $\{2,1,1,1\}$ & $\bm{0},\mathsmaller{1},\mathsmaller{2}$ \\
 & 2 & $\{2,2,1,1\}$ & $\bm{0},\mathsmaller{1},\mathsmaller{2}$ \\
 & 3 &  $\{2,2,2,1\}$ & $\bm{0},\mathsmaller{1}$\\
 \midrule
$\text{KS}_\kappa$ & 1  & $\{1.5,1.5,1,1\}$
& $\bm{0}$\\ 
 & 2 &$\{1.5,1.5,1.5,1.5\}$
& $\bm{0}$\\ 
 & 3 & $\{2,2,1.5,1.5\}$
& $\bm{0}$\\ 
 \midrule
$\text{CDW}$ & 2  & $\{1.5,1.5,1.5,1.5\}$
& $\bm{0}$\\ 
 & 3 & $\{2,2,1.5,1.5\}$
& $\bm{0},\bm{2}$
\\\bottomrule[1.3pt]
\end{tabular}
\caption{\textbf{Properties of phases at finite displacement field.} The phases listed here appear in the finite displacement field phase diagrams of Fig.~\ref{fig:U_theta_phase_diagram}. $\widetilde{\text{FB}}$ represents the family that is obtained from the strong-coupling family $\text{FB}$ via a displacement field-tuned topological transition. $\{\text{flavor occ.}\}$ lists the flavor occupations in descending order. The Kekul\'e spiral (KS$_\kappa$) and charge density wave (CDW) can have fractional occupations due to intervalley coherence or translation symmetry breaking. $C$ denotes the possible Chern numbers, with large bold entries corresponding to the dominant ground-state HF phases obtained in Fig.~\ref{fig:U_theta_phase_diagram}.\label{tab:finite_U_phases}}
\end{table}

Fig.~\ref{fig:BM_finiteU} illustrates the evolution of the non-interacting band structure as a function of the interlayer potential $U$. The overall bandwidth widens as $U$ increases, but the most significant changes occur at the mBZ corners. At zero displacement field, the bands are nearly degenerate at $E=0$ around the moir\'e minivalleys (Fig.~\ref{fig:BM_zeroU}a), but split into sublattice polarized bands for finite $U$. While the $A$ sublattice remains close to $E=0$, the $B$ sublattice shifts significantly in energy with opposite signs at $\kappa$ and $\kappa'$. This is because the $B$ bands carry a significant momentum-contrasting layer dipole moment. The color plots in Fig.~\ref{fig:BM_finiteU} show that the previously isolated high-energy lobes in the mBZ merge into a single `fidget-spinner' feature centered around $\kappa$ ($\kappa'$) in valley $K$ ($\bar{K}$) for large $U$.

\begin{figure}[t]
    \includegraphics[width=0.9\columnwidth]{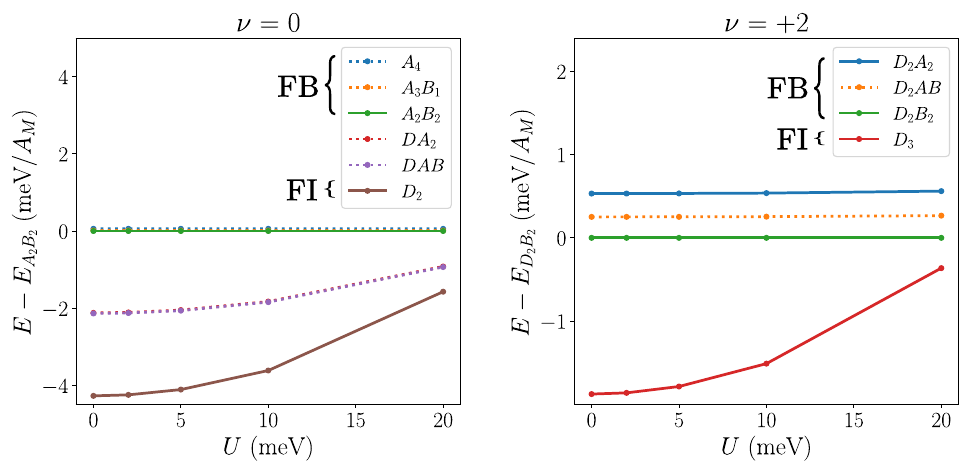}
    \caption{\textbf{Energetic competition between strong-coupling phases at finite displacement field.} Self-consistent HF energies of different strong-coupling classes. Labelling is the same as Fig.~\ref{fig:strong_coupling_energies}. System size is $12\times 12$, $\theta=1.8^\circ$ and $w_{\text{AA}}=70\,\text{meV}$.}
    \label{fig:strong_coupling_energies_U}
\end{figure}

The HF phase diagrams as a function of interlayer potential and twist angle are presented in Fig.~\ref{fig:U_theta_phase_diagram}, which show mostly gapped states. At $U=0$, we recover the strong-coupling phases discussed in Sec.~\ref{sec:phase_wAA}. As demonstrated in Fig.~\ref{fig:strong_coupling_energies_U}, the delicate competition between strong-coupling classes in the same family (defined as sharing the same flavor occupations modulo flavor symmetries) persists as $U$ is ramped up. As the interlayer potential broadens the bandwidth, it generally favors strong-coupling families with fewer fully-occupied flavors, i.e.~the FB family. This is consistent with the narrowing of the FI region in Fig.~\ref{fig:U_theta_phase_diagram} as $U$ increases. Beyond a threshold value of $U$, which is comparable across the filling factors, the phase diagram contains phases which cannot be understood as simple strong-coupling insulators. The flavor occupations and Chern numbers of these new phases are summarized in Tab.~\ref{tab:finite_U_phases}.

\begin{figure}[t]
    \includegraphics[width=1\columnwidth]{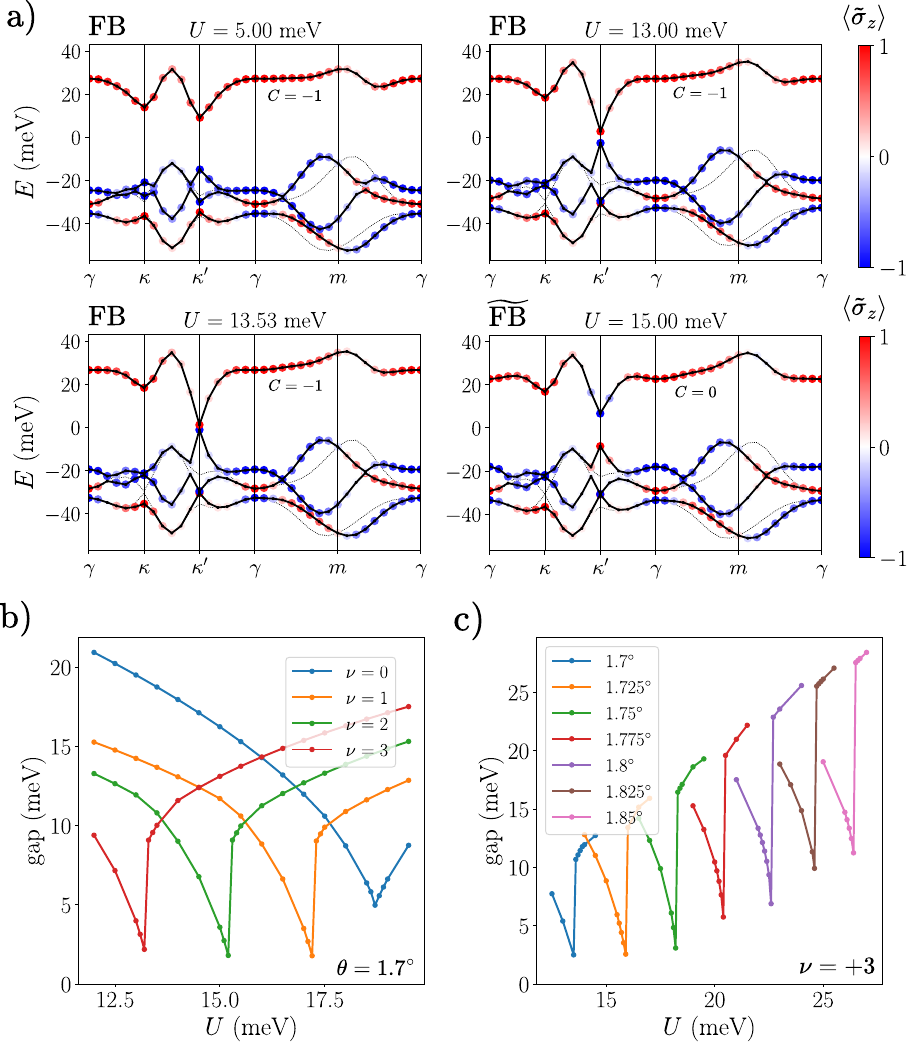}
    \caption{\textbf{Displacement field tuned topological transition.} a) First three plots show the HF band structures for the $D_3B$ phase at $\nu=+3$ (see Fig.~\ref{fig:strong_coupling_nu3}a for a schematic) for increasing interlayer potential $U$. For sufficiently large $U$, the system undergoes a topological transition in valley $\bar{K}$ to a phase where conduction band no longer has a non-zero Chern number $C$. Colored dots indicate the Chern-sublattice polarization. Only the bands in the partially occupied spin sector are shown. System size is $18\times 18$, $\theta=1.7^\circ$, and $w_{\text{AA}}=75\,\text{meV}$. b) Gap versus $U$ for different fillings $\nu$ at $\theta=1.7^\circ$. c) Gap versus $U$ for different twist angles at $\nu=+3$. }
    \label{fig:topological_transition_nu3}
\end{figure}

For most values of $\theta$, the first non-strong-coupling phase that is encountered as $U$ increases is the $\widetilde{\text{FB}}$ family. The HF solution in this region shares the same symmetries and similar Chern basis occupations as the neighboring FB phase at smaller $U$. When entering the transition from the FB phase (which is restricted to $\theta\lesssim1.8^\circ$ for $\nu\neq 3$), the gap is greatly suppressed, implying a continuous or weakly first-order transition (which is the case within our HF calculations). However, the $\widetilde{\text{FB}}$ family possesses a distinct set of possible Chern numbers, which can be seen by comparing Tab.~\ref{tab:strong_coupling} and ~\ref{tab:finite_U_phases}. 

Fig.~\ref{fig:topological_transition_nu3}a reveals that this arises from a topological phase transition at the mBZ corners, using a state in the $[D_3B]$ class at $\nu=+3$ as an example. In this calculation, the empty conduction band is primarily composed of the $(\bar{K},\downarrow,A)$ Chern band. As $U$ increases, the energy of the $B$ sublattice in the valence band for the $\bar{K},\downarrow$ flavor sector rapidly increases and closes the gap at $\kappa'$. Across the topological transition, the bands get inverted so that the conduction band becomes topologically trivial, but the bands largely retain their original flavor and sublattice polarized character elsewhere in momentum space. Fig.~\ref{fig:topological_transition_nu3}b demonstrates that the gap at the band closing point is sharply suppressed for the other fillings as well, with the threshold field decreasing slightly with density. 
The possible Chern numbers of the $\widetilde{\text{FB}}$ family in Tab.~\ref{tab:finite_U_phases} are obtained by using a new effective set of sublattice Chern numbers $C_{K,s,A}=0, C_{K,s,B}=-1, C_{\bar{K},s,A}=0, C_{\bar{K},s,B}=1.$, c.f.~Eq.~\ref{eq:Chern_numbers}.
Fig.~\ref{fig:topological_transition_nu3}c shows that the gap minimum is reduced for smaller twist angles. Hence, the system realizes a set of displacement field-tuned topological transitions, which we emphasize occur in the strongly-interacting regime where there is still significant generalized flavor-sublattice ferromagnetism. We comment that the asymmetry and discontinuity of the HF gap about the transition point is similar to that seen in studies of the inverted charge transfer mechanism relevant for transition metal dichalcogenide heterobilayers~\cite{devakulinverted2022}, which also realizes a topological band inversion in the strongly interacting regime. 

As in the low-$U$ regime, there is a close energetic competition in the $\widetilde{\text{FB}}$ phase between classes which share the same flavor occupation numbers but differ in the sublattice polarizations. For larger $U$, these states can also become unstable to a weak-coupling excitonic instability. 

Experimentally, the topological transitions would manifest as a dip in the resistive peak or incompressibility as a function of displacement field. 
Furthermore, since the phases below/above the transition have generically different Chern numbers, another signature would be a change in the anomalous Hall effect at the transition, as well as differing slopes of various features in the $\nu-B$ plane.

\begin{figure}[t]
    \includegraphics[width=1\columnwidth]{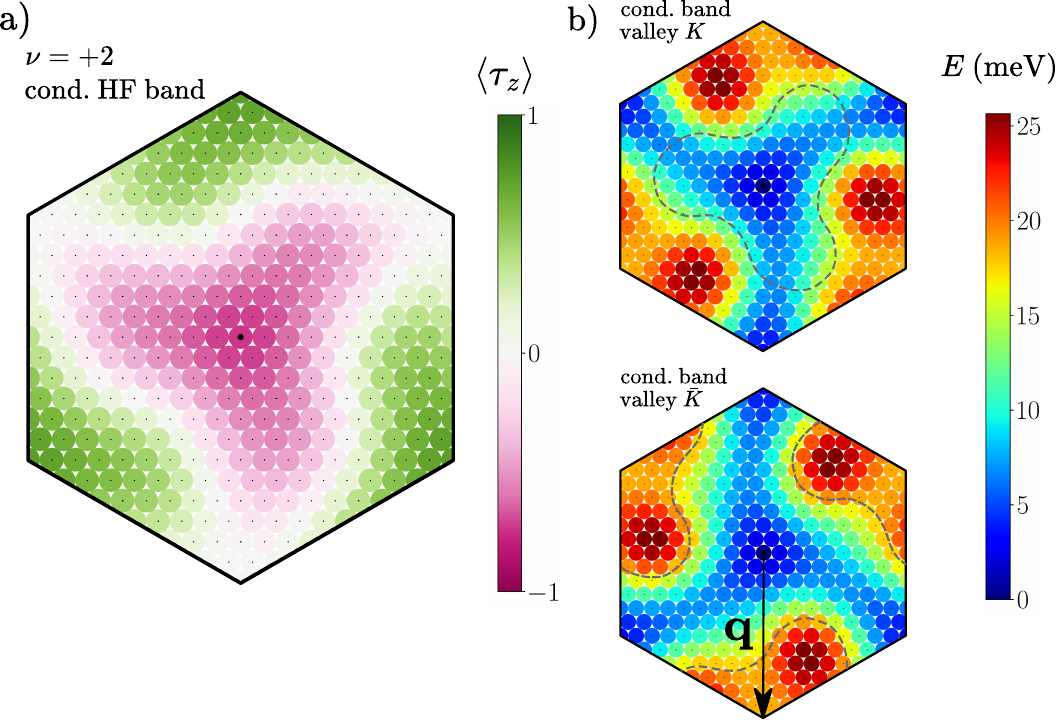}
    \caption{\textbf{Kekul\'e spiral (KS$_\kappa$) phase at finite displacement field.} a) Valley polarization $\langle\tau_z\rangle$ for the HF conduction band at $\nu=+2$ in one spin sector (the system is $SU_S(2)$-symmetric). The momentum is measured using the original mBZ coordinates in valley $K$ ($\tau_z=+1$). b) Non-interacting dispersion of the BM conduction bands in both valleys. Dashed contour in valley $\bar{K}$ roughly indicates the high-energy lobe there. The same contour is shown in $K$, but shifted by $-\bm{q}$, where $\bm{q}$ is the Kekul\'e spiral wavevector. In the KS$_\kappa$ phase, any intervalley coherence hybridizes a momentum $\bm{k}$ in $K$ with $\bm{k}+\bm{q}$ in $\bar{K}$. The $\kappa$ subscript indicates that $\bm{q}$ lies at one of the $C_3$-symmetric momenta $\kappa$ or $\kappa'$. System size is $18\times 18$, $\theta=1.8^\circ,w_{\text{AA}}=75\,\text{meV},U=50\,\text{meV}$.}
    \label{fig:Kek_k}
\end{figure}

For all non-zero integer fillings, the system enters the Kekul\'e spiral (KS$_\kappa$) phase for sufficiently large interlayer potentials. This state is closely related to the incommensurate Kekul\'e spiral (IKS) order which has been theoretically proposed~\cite{kwan2021kekule,wagner2022global,wang2022kekul} and experimentally observed in TBG~\cite{nuckolls2023quantum} and mirror-symmetric trilayer graphene~\cite{kim2023imaging}. The KS$_\kappa$ state preserves TRS $\hat{\mathcal{T}}$ but breaks moir\'e translation symmetry $\hat{T}_{\bm{a}_i}$ and valley $U(1)_v$. The flavor occupations (Tab.~\ref{tab:finite_U_phases}) and lack of significant polarization in the Chern-sublattice further distinguishes this state from the strong-coupling phases or their descendants obtained via excitonic instabilities or topological transitions. However, the KS$_\kappa$ state preserves a twisted translation symmetry $\hat{T}'_{\bm{a}_i}=\hat{T}_{\bm{a}_i}e^{-i\frac{\bm{q}\cdot\bm{a}_i}{2}\tau_z}$ which includes a valley rotation. This property derives from the structure of the density matrix in the intervalley channel, where $P_{\bm{k}K;\bm{k}'\bar{K}}$ is forced to be zero unless $\bm{k}'=\bm{k}+\bm{q}$. The stabilization of this phase is traced to the momentum dependence of the dispersion for large $U$. By `boosting' the $\bar{K}$ valley by the Kekul\'e spiral wavevector $\bm{q}$ (Fig.~\ref{fig:Kek_k}b), the low and high energy features in the two valleys can be superposed according to the `lobe' principle described in Ref.~\cite{kwan2021kekule}. As shown in Fig.~\ref{fig:Kek_k}, the shape of the non-interacting bands is imprinted on the resulting valley polarization of the HF bands. In contrast to the $\hat{C}_{2z}$-symmetric IKS which was originally proposed in the context of $\hat{C}_{3z}$-breaking heterostrain, the wavevector of the KS$_\kappa$ appears to be pinned to one of the $\hat{C}_{3z}$-symmetric corners of the mBZ, and the lobes have a sizable sublattice polarization. 
The KS$_\kappa$ state is difficult to experimentally discriminate from the $C=0$ members of the $\widetilde{\text{FB}}$ phase. The most direct test would be observation of significant correlation-induced Kekul\'e distortion on the graphene scale using STM~\cite{nuckolls2023quantum,kim2023imaging}, since none of the other candidate phases have an appreciable amount of intervalley coherence.

Finally at larger angles, there is a small sliver of the phase diagram which consists of a charge density wave (CDW) on the moir\'e scale that breaks TRS even in the absence of valley polarization. The TSB occurs predominantly in the $B$ sublattice which takes advantage of the dispersive momentum features by folding the bands. This quadruples the unit cell area because the periodicity along both moir\'e axes is doubled. At $\nu=+2$, the sublattice basis occupations are consistent with starting from an $[A_4]$ state and occupying half a $B$ band in each flavor. 

\section{Discussion}\label{sec:discussion}

At low displacement fields, as well as intermediate fields (the $\widetilde{\text{FB}}$ phase), our findings paint a picture of a multitude of closely competing (near)-strong-coupling states with contrasting flavor and sublattice polarizations, and varying electronic topology. A key question is how this manifold is ultimately split, which is relevant for resolving the low-temperature physics. Different strong-coupling classes within the same family (see Fig.~\ref{fig:strong_coupling_schematic} for schematic of labelling of strong-coupling states) have similar energies within $1~\text{meV}$ per moir\'e cell as illustrated in Fig.~\ref{fig:strong_coupling_energies} and anticipated from strong-coupling theory (Sec.~\ref{sec:strong_coupling}). While the HF calculations and strong-coupling analysis show a consistent preference towards maximizing the occupation of the $B$ bands for $\nu\geq 0$, the splittings are small enough that they could be reversed by effects not captured in our modelling like residual coupling to the hBN substrate. Within a given strong-coupling class, the remaining choice of the state pertains to the flavor degrees of freedom. As an example, consider the $[D_2B_2]$ class of the FB family at $\nu=+2$ --- a particular state can be chosen by specifying the valley and spin quantum numbers of the two unoccupied $A$ bands. Some degeneracies are expected to remain exact, such as the global $SU(2)_S$ spin-symmetry or spinless time-reversal $\hat{\mathcal{T}}$, unless they are deliberately broken with e.g.~an external magnetic field. Others are only exact because certain terms have been neglected from the Hamiltonian in our study. These include various Hunds couplings which are not invariant under independent spin rotations in the two valleys. If we assign one empty $A$ band to each valley, then their spins will align (anti-align) if the correction is ferromagnetic (anti-ferromagnetic). The sign of the Hunds term involves opposite contributions from optical phonons and intervalley Coulomb scattering, and is difficult to pin down theoretically~\cite{chatterjeeSymmetryBreakingSkyrmionic2020}, though there is experimental evidence that this is anti-ferromagnetic in TBG~\cite{morissette2022electron}. 

As discussed in detail in Sec.~\ref{sec:strong_coupling}, there is another type of degeneracy which is unique to the mean-field nature of Hartree-Fock, and corresponds to acting with the spinless time-reversal operation only for one spin projection. For the $[D_2B_2]$ class at $\nu=+2$, a scenario where this applies is where the empty $A$ bands have flavors $(K,\uparrow)$ and $(\bar{K},\downarrow)$, versus $(K,\uparrow)$ and $(K,\downarrow)$. Crucially, these two choices have $|C|=0$ and $2$ respectively, but are degenerate in our calculations since the HF Hamiltonian is quadratic. Furthermore, this degeneracy is also not split to lowest order in the strong-coupling perturbation theory of Sec~\ref{sec:strong_coupling} .  However, `time-reversal in one spin projection' cannot be an exact symmetry since it is neither unitary nor anti-unitary. Therefore, quantum fluctuations which introduce deviations from a single Slater determinant will split this degeneracy. These `Hartree-Fock symmetries' also occur to a limited extent in other moir\'e systems like TBG~\cite{kwan2021kekule,kwan2023electronphonon}, so an interesting future direction is to systematically investigate how the corresponding degeneracies are lifted.

The small energy differences between different classes and families of strong-coupling states can be traced to the perturbative proximity to the chiral-flat strong-coupling limit, where all generalized ferromagnets are split only by a small amount corresponding to the sublattice polarization (Eq.~\ref{eq:main_subsplit}). In TBG, this close competition is sidestepped in many devices by the presence of heterostrain~\cite{kerelsky2019maximized,choi2019electronic,xieSpectroscopicSignaturesManybody2019,mespleheterostrain2021}, which allows the IKS to undercut the strong-coupling manifold~\cite{kwan2021kekule,wagner2022global,wang2022kekul,nuckolls2023quantum,kim2023imaging}. It would be useful to investigate the effect that strain has on the band structure~\cite{huder2018electronic,bi_designing_2019} and phase diagram of h-HTG. We anticipate that it is less susceptible to strain-induced IKS order, since the larger twist angle enhances the interaction scale, and the homogeneous charge density reduces the tendency to form significant momentum-dependent features in the interacting band structure. Owing to the super-moir\'e structure, strain could also influence or be absorbed into the relaxation of domains in a non-trivial fashion~\cite{devakul2023magicangle}. We note though that the related KS$_\kappa$ state appears to emerge for moderate displacement fields already in the absence of strain (Fig.~\ref{fig:U_theta_phase_diagram}) --- it would be interesting to check whether this persists with more sophisticated numerical techniques like DMRG~\cite{soejimaEfficientSimulationMoire2020,wang2022kekul}.

We restrict our numerical calculations to non-negative integer fillings due to the exact particle-hole symmetry we impose on the model. However, it is known that in other systems for which particle-hole symmetry is commonly assumed theoretically, the experimental data show pronounced particle-hole asymmetry in fundamental observables like the positions of the dominant correlated insulators and superconducting domes (see for instance Ref.~\cite{saito2020independent}). We anticipate that similar considerations will apply to our system, and that refinements to the theoretical modelling that aim to cure this deficiency, e.g.~by adding terms to the continuum model~\cite{carr2019exact,fang2019angledependent,guinea2019continuum,kangPseudomagneticFieldsParticlehole2022,vafekcontinuum2023}, could be applied to h-HTG.

While we have focused on the integer phase diagrams, our results influence the physics at non-integer fillings. At low dopings away integer $\nu$, the Fermi surfaces are likely controlled by the interacting band structure of the parent insulator. Information on the number of Fermi surfaces and their sizes is invaluable since it can be extracted via measurements of the Landau fans and their degeneracies (though quantum oscillations may be hard to detect if the effective masses are too large). Consider for instance electron-doping the particular $\nu=+2$ $[D_3]$ state shown in Fig.~\ref{fig:strong_coupling_nu2}b at zero displacement field. The electrons initially form three $\hat{C}_3$-related Fermi surfaces, and are predominantly of $B$ character. Note that these conclusions may be altered in the presence of extrinsic $\hat{C}_{3z}$-breaking strain, or nematicity induced by polarizing in momentum space~\cite{dongpolarized2023}. For other states, as discussed in Sec.~\ref{sec:phase_wAA} for electron-doping the $\nu=+3$ $[D_3 B]$ insulator, the system may additionally undergo a finite-filling sublattice transition where some of the carriers abruptly switch from one sublattice to the other. This would truncate the Landau fans emanating from the parent integer, and lead to filling-dependent modulations in the spectral function in STM/STS~\cite{wongCascadeElectronicTransitions2020} or the electronic compressibility in SET~\cite{zondinerCascadePhaseTransitions2020} measurements. Pinning down the precise pattern of flavor/sublattice transitions and their corresponding signatures would require the HF computations to be extended to all non-integer fillings. Given that the $B$ bands have larger dispersive features, we expect that a common driving force behind the sublattice transitions is a minimization of the kinetic energy of the $B$ quasiparticles. 

At fractional non-integer fillings, the system can form correlated insulators beyond simple flavor-symmetry-broken Fermi liquids. Accessible within HF are translation-symmetry-breaking (TSB) phases obtained by folding the mBZ and inducing a moir\'e charge density wave. (Note that a modified Lieb-Schulz-Mattis theorem forbids a pure flavor spiral order from being gapped at non-integer filling~\cite{kwan2021kekule}.) As the $B$ bands have significant momentum-dependent features, we expect the TSB order parameter to be concentrated here, as for the integer CDW phase in Sec.~\ref{fig:U_theta_phase_diagram}. Preliminary calculations show that this is indeed the case, and find various TSB insulators at various third- and half-fillings. We defer a detailed exploration of such phases to future work. As proposed in \cite{devakul2023magicangle}, another class of candidate states is fractional Chern insulators at, say, $\nu=3+\frac{1}{3}$ or $3+\frac{2}{3}$. This scenario is motivated by from the narrow quasiparticle dispersion, energetic isolation, and ideal quantum geometry~\cite{parameswaranFractionalQuantumHall2013,liuFractionalChernInsulators2012,royBandGeometryFractional2014,jacksonGeometricStabilityTopological2015,parkerFieldtunedZerofieldFractional2021,ledwith2022vortexability,ledwithFamilyIdealChern2022,dong2022exact,wangExactLandauLevel2021,ledwithStrongCouplingTheory2021,gao2022untwisting,varjas2022} of the partially occupied $A$ band (Fig.~\ref{fig:strong_coupling_nu3}), and would need to checked by DMRG~\cite{parkerFieldtunedZerofieldFractional2021} or exact diagonalization~\cite{abouelkomsanParticleHoleDualityEmergent2020,repellinChernBandsTwisted2020,wilhelmInterplayFractionalChern2021}. The modelling is simplified by the substantial suppression of interaction renormalization. Interesting correlated states have also been proposed for higher $|C|=2$ bands~\cite{dong2022exact,wang2022origin,wuBlochModelWave2013,Kumar2014Generalizing,barkeshliTopologicalNematicStates2012,barkeshliTwistDefectsProjective2013,wilhelm2023noncoplanar,polshyn2020electrical,behrmannModelFractionalChern2016,wangFractionalQuantumHall2012,liuFractionalChernInsulators2012,trescherFlatBandsHigher2012,yangTopologicalFlatBand2012,sterdyniakSeriesAbelianNonAbelian2013,andrewsStabilityFractionalChern2018,andrewsStabilityPhaseTransitions2021}, which could be relevant if the energetics at fractional fillings prefer that $B$ bands are partially filled instead.

We expect that real samples of HTG will form a super-moir\'e pattern that locally relaxes into domains of h-HTG and its $\hat{C}_{2z}$-related partner $\bar{\text{h}}$-HTG~\cite{devakul2023magicangle}, as illustrated in Fig~\ref{fig:D_nu_splash}. 
The structure, as a whole, therefore has $\hat{C}_{2z}$ symmetry on the super-moir\'e scale, despite the symmetry being absent within the h-HTG structure.
The Chern numbers of the sublattice basis in $\bar{\text{h}}$-HTG are obtained by taking $A\leftrightarrow B$ and $K\leftrightarrow \bar{K}$ in Eq.~\ref{eq:Chern_numbers}. 
Since the central bands in the two domains carry opposite valley Hall numbers, the domain walls induce a network of gapless topological edge modes that cross the remote band gap at $|\nu|=4$, which can be traced via local imaging. 
The shape of this network is triangular in the pristine limit, but may deform due to factors such as twist angle disorder and strain. 
For other integer fillings, the correlated insulators in the h-HTG and $\bar{\text{h}}$-HTG will have experimental signatures in the form of resistance peaks, as well as signatures of the displacement-field tuned topological transition.

Furthermore, there is additional physics arising from the choice of correlated state in two adjacent domains~\cite{kwan2021domain,grover2022chern}. Degeneracies within each domain can be split by the configuration in neighboring domains. As a concrete example, we consider the strong-coupling $[D_3B]$ class in h-HTG and the equivalent $[D_3A]$ class in $\bar{\text{h}}$-HTG at $\nu=+3$, which have $C=\pm 1$.  Ignoring spin for simplicity, there a freedom in assigning valleys to the unoccupied band in each domain (see schematic in Fig.~\ref{fig:strong_coupling_nu3}a). For equal (opposite) valleys, the domains have opposite (equal) Chern numbers, leading to copropagating (counterpropagating) edge modes at the interface, and realizing Chern (valley) domain walls. 
While a detailed analysis of domain wall energetics is an important topic for future work, we now argue that Chern domain walls should be favored at zero magnetic field.
Relaxation calculations show that the aperiodic interface channels are only a few moir\'e lengths wide~\cite{devakul2023magicangle}, suggesting that mutual exchange physics of the two domains, which is not possible between opposite valleys, is important. 
Furthermore, intervalley coherence is suppressed due to the mismatch of Chern numbers in the two valleys~\cite{bultinckMechanismAnomalousHall2020}, which discourages spatial texturing of the valley wall. 
Hence, we anticipate that Chern domains will be energetically favored,
thus resulting in a ``Chern mosaic''~\cite{grover2022chern} of h-HTG and $\bar{\text{h}}$-HTG domains carrying opposite Chern numbers.
This possibility can be numerically tested with similar techniques as Ref.~\cite{kwan2021domain} and visualized using a SQUID-on-tip~\cite{tschirhart2021imaging,grover2022chern}. 
A sufficiently strong perpendicular magnetic field may counteract this and induce valley domain walls, since it couples to the orbital magnetization of the domains. 

Another important question is the consequence of such Chern domains physics in transport.  
Even when the h-HTG and $\bar{\text{h}}$-HTG domains individually realize Chern insulators, thus breaking $\mathcal{T}$, it is possible that $\hat{C}_{2z}\mathcal{T}$ is restored at the super-moir\'e scale, as is the case for the above Chern mosaic.
We remark that this super-moir\'e scale $\hat{C}_{2z}\mathcal{T}$ symmetric Chern mosaic is unique to HTG, and is absent in hBN-aligned TBG~\cite{grover2022chern,shi2021moire}.
The presence of $\hat{C}_{2z}\mathcal{T}$ forbids a non-zero net Hall conductivity in the thermodynamic limit.
However, this vanishing Hall conductivity relies on the precise cancellation of currents in h-HTG and $\bar{\text{h}}$-HTG domains, which are potentially several hundred nanometers wide and spatially separated.
Because of this, extrinsic effects in real mesoscopic devices, such as various forms of strain or twist angle disorder (which can vary greatly on the micron scale), the precise placement of contacts with respect to the domains, or boundary effects on the domain sizes and shapes, likely mean that the super-moir\'e scale $\hat{C}_{2z}\mathcal{T}$ symmetry is not relevant to electronic transport properties at experimentally relevant scales.
Hence, we expect a remnant (non-quantized) anomalous Hall effect to be experimentally measurable even in the Chern mosaic state, reflecting the broken $\mathcal{T}$ symmetry.
Experimental determination of the Chern numbers can be achieved using the Streda formula in a finite magnetic field.
Local techniques leveraging an STM~\cite{wongCascadeElectronicTransitions2020,choi2021correlation,choi2021interactiondriven,nuckolls2023quantum,kim2023imaging,turkel2022orderly,nuckollsStronglyCorrelatedChern2020,zhang2023local} and SET~\cite{xieFractionalChernInsulators2021,yu2022correlated,yu2022spin} are able resolve this information within each domain.

In conclusion, we have presented a comprehensive analysis of the interacting physics at integer fillings of HTG using complementary methods of strong-coupling theory and HF.  
Our analysis reveals h-HTG as an ideal platform for realizing strong-coupling physics, with interactions dominating over bandwidth at all integer fillings.
We uncover a rich heirarchy of correlated insulating states, many of which are topological, and predict topological phase transitions as a function of displacement field.
We discuss in detail the experimental ramifications of our findings, which can be readily tested with existing experimental techniques.
Our work paints a rich picture of moir\'e-scale interaction-driven topology intertwined with the super-moir\'e scale topological domains, paving the way for future studies of interacting physics in HTG.

\begin{acknowledgments}
We thank Ashvin Vishwanath and Ben Feldman for valuable comments on the manuscript. 
TD and YK thank Liqiao Xia, Aviram Uri, Sergio de la Barrera, Ziyan Zhu, Liang Fu, and Pablo Jarillo-Herrero for valuable discussions and collaboration on related projects.
YK thanks Glenn Wagner for previous collaboration on the Hartree-Fock code. PL thanks Eslam Khalaf and Ashvin Vishwanath for valuable discussions and previous collaboration on related projects.
\end{acknowledgments}

%


\newpage

\onecolumngrid

\begin{appendix}

\section{Strong Coupling Perturbation Theory}\label{app:strong_coupling}

In this appendix we analyze the splitting of strong coupling states in h-HTG upon moving away from the chiral, dispersion-free limit where they are all degenerate. Before doing so it is useful to review some notation.
We will use the ``hatted'' notation for second quantized operators $\hat{A}$ in terms of their first quantized versions $A$. Hatting preserves commutators:
\begin{equation}
  \hat{A} = \sum_{I,J} c^\dag_I A_{IJ} c_J, \qquad \hat{[A,B]} = [\hat{A},\hat{B}].
  \label{eq:hatteddefn_supp}
\end{equation}
Here, $I,J$ label all states in the band-projected single-particle Hilbert space: they are multi-indices that combine both momentum $\bk$, spin, valley, and sublattice. For example, $\hat{\rho_\bq} = \sum_{I J} c^\dag_I \rho_{\bq IJ} c_J$ is the second quantized version of the first quantized $\rho_\bq$ where 
\begin{equation}
  \rho_{\bq IJ} = (\rho_{\bq})_{\bk \bk'}^{\tau \tau', \tilde{\sigma} \tilde{\sigma}'} = \Lambda_{\bq}^{\tau, \tilde{\sigma} \tilde{\sigma}'}(\bk) \delta_{\bk + \bq, \bk'} \delta_{\tau \tau'} \delta_{s s'}, \qquad \Lambda^{\tilde{\sigma} \tilde{\sigma}'}_{\bq}(\bk) = \braket{u_{\bk \tilde{\sigma}}^\tau}{u_{\bk +\bq \tilde{\sigma}'}^\tau}.
  \label{eq:densityexample}
\end{equation}
Here $\tau = \pm = K,\bar{K}$ labels the graphene valley, $\tilde{\sigma} = \pm = A,B$ labels the sublattice band, and $s = \pm = \uparrow, \downarrow$ labels the spin. For the appendices, from now on, we we will drop the tildes on the sublattice band labels and Pauli matrices for convenience: we will not have to refer to the microscopic sublattice.
Since unperturbed strong coupling states are exact Slater determinants, we will be interested in expectation values of Slater determinant states in this section. These may be evaluated by Wick's theorem
\begin{equation}
  \langle \hat{A} \rangle =  \Tr P A, \qquad \langle \hat{A} \hat{B} \rangle = \Tr PAB - \Tr APBP + \Tr AP \Tr BP = \frac{1}{2} \Tr [A,P][P,B] + \Tr AP \Tr BP,
  \label{eq:expectationvalues}
\end{equation}
where $P_{IJ} = \langle c^\dag_J c_I \rangle$ is the Hartree Fock projector: $P^2=P$. Here, we are using a capital $\Tr$ to denote traces over the multiindices $I,J$. For traces over band indices, that don't include sums over momenta, we use $\tr$. 

The full Hamiltonian is
\begin{equation}
  \H = \hat{h} + \frac{1}{2A} \sum_\bq V_\bq \delta \hat{\rho}_\bq  \delta \hat{\rho}_{-\bq}, 
  \label{eq:fullHam}
\end{equation}
where $A = N_M A_M$ is the sample area in terms of the number of unit cells $N_M$ and the unit cell area $A_M$. We have used
\begin{equation}
  \delta \hat{\rho}_\bq = \hat{\rho}_\bq - 4\ov{\rho}_\bq, \qquad \overline{\rho}_\bq = \frac{1}{4} \sum_{\tau \sigma} \overline{\rho}^{\tau \sigma}_\bq, \qquad \overline{\rho}^{\tau \sigma}_\bq =  \sum_\bG  \delta_{\bq,\bG}  \Lambda^{\tau, \sigma \sigma }_\bG(\bk)
\end{equation}
which is the density measured relative to half filling of the flat bands in a periodic gauge $c_\bk = c_{\bk+\bG}$. The background density $-4\ov{\rho}_\bq$, where $\ov{\rho}_\bq$ is the average density of a fully filled band, is only nonzero if $\bq$ is a reciprocal lattice vector, by translation symmetry. The factor of four arises because at charge neutrality four out of eight bands are filled.

\subsection{Symmetric, Dispersion-Free, Limit}

We begin with the idealized limit of symmetric form factors, $[\Lambda_\bq(\bk),\sigma_z] = 0$ from chiral symmetry, such that $\Lambda_\bq^{\tau, \sigma \sigma'}(\bk) = \Lambda^{\tau, \sigma}_\bq(\bk) \delta_{\sigma \sigma'}$ and zero dispersion $\hat{h}$. Then, the form factors $\Lambda_\bq(\bk)$ are $8 \times 8$ diagonal matrices, with diagonal elements $\Lambda_\bq^{\tau \sigma}(\bk)$ (we leave the spin index implicit, since $\Lambda$ is spin-independent). Note that we are only neglecting the sublattice off-diagonal part of the form factor here; we include the $\kappa = w_{AA}/w_{AB} > 0$ contributions to the diagonal form factor in practice. 
In this idealized limit the form factors, and the interacting Hamiltonian, have a $U(2) \times U(2) \times U(2) \times U(2)$ symmetry consisting of spin and charge rotations within each sublattice and valley sector.

Since the form factor is diagonal in sublattice and valley, we will shortly see that all many-body states $\ket{\Psi}$ that consist of completely filling some number of sublattice and valley diagonal bands are exact eigenstates. Such states are described by translationally symmetric, $\bk$-independent, Hartree-Fock projectors $P$ that are flavor and sublattice diagonal. We will use the notation
\begin{equation}
    P = \sum_{\tau \sigma} P^\tau_\sigma \frac{1 + \tau \tau_z}{2} \frac{1 + \sigma \sigma_z}{2}, \qquad P_\sigma = \sum_\tau P^\tau_\sigma \frac{1 + \tau \tau_z}{2}, \qquad P^\tau = \sum_\sigma P^\tau_\sigma \frac{1 + \sigma \sigma_z}{2}
    \label{eq:unperturbedstates}
\end{equation}
and
\begin{equation}
    P_\sigma = \frac{1}{2}(1+Q_\sigma), \quad P^{\tau} = \frac{1}{2}(1 + Q^\tau), \quad P_{\sigma}^\tau = \frac{1}{2}(1 + Q_\sigma^\tau).
\end{equation}
The matrix $Q^{\tau}_\sigma$ is a $2 \times 2$ matrix that describes the spin-occupations in valley $\tau$ and sublattice $\sigma$. It is acted on by the $U(2)$ spin and charge rotation associated to this spin and valley: $Q^{\tau}_\sigma \to (U_\sigma^\tau)^\dag Q^\tau_\sigma U_\sigma^\tau$, though the sublattice and valley occupations 
\begin{equation}
\nu^{\tau}_\sigma = \frac{1}{2} \tr Q^\tau_\sigma
\end{equation}
are invariant under this action.

The states described by \eqref{eq:unperturbedstates} are distinguished by the fact that they commute with $\Lambda_\bq(\bk)$, so that the density operator cannot scatter occupied states to deoccupied states. To see that these states are exact eigenstates, we note that $\hat{\rho}_\bq \ket{\Psi} = 0$ unless $\bq = \bG$ is a reciprocal lattice vector, since all other scattering is Pauli blocked. For reciprocal lattice wavevectors, we have
\begin{equation}
  \delta \hat{\rho}_\bG \ket{\Psi} = \sum_{\bk \tau \sigma} \Lambda_\bG(\bk)\left(c^\dag_{\bk \tau \sigma}  c_{\bk \tau \sigma} - \frac{1}{2} \right) \ket{\Psi} = \sum_{\sigma \tau} \nu_{\sigma}^\tau \rho_{\bG}^{\tau \sigma}  \ket{\Psi}
  \label{eq:densityeval}
\end{equation}
Since the only opertor in the disperisonless Hamiltonian is the density operator, \eqref{eq:densityeval} implies that $\ket{\Psi}$ is an eigenstate of the Hamiltonian. We now assess which states have the smallest energy and what this energy depends on.

The form factors for different sublattices are not symmetry related since the associated bands have different $\abs{C}$, but the form factors for different valleys are related by time reversal: $\Lambda^{(-\tau) \sigma}_{\bq}(\bk) = \overline{\Lambda^{\tau \sigma}_{-\bq}(-\bk)} = \Lambda^{\tau \sigma}_{\bq}(-\bk - \bq)$. This implies that the background density is the same for the bands in each valley related by time reversal, as one would expect: $\ov{\rho}_\bG^{(-\tau) \sigma} = \ov{\rho}_\bG^{\tau \sigma} = \ov{\rho}_\bG^\sigma$. We also now have $\ov{\rho}_\bG = \frac{1}{2}\sum_{\sigma = \pm} \ov{\rho}^{\sigma}_\bG$. Thus, only the sublattice polarization affects the eigenvalue $\eqref{eq:densityeval}$. In particular we have
\begin{equation}
  \delta \rho_\bG \ket{\Psi} = \left(\ov{\rho}_\bG \nu + \ov{\rho}^z_\bG \nu_z\right) \ket{\Psi}
  \label{eq:subpoleval}
\end{equation}
where we have defined $\ov{\rho}^z_\bG = \frac{1}{2}\sum_{\sigma = \pm} \sigma \ov{\rho}_\bG^\sigma$, the filling $\nu = \frac{1}{2} \tr Q$, and sublattice polarization $\nu_z = \frac{1}{2}\tr Q \sigma_z$. The total energy is then
\begin{equation}
  \H \ket{\Psi} = E_0 \ket{\Psi}, \qquad E_0 = \frac{1}{2A} \sum_\bG V_\bG (\ov{\rho}_\bG \nu + \ov{\rho}^z_\bG \nu_z)(\ov{\rho}_{-\bG} \nu + \ov{\rho}^z_{-\bG} \nu_z).
  \label{eq:totalenergy}
\end{equation}

At charge neutrality, $\nu = 0$, we see that any state that has zero sublattice polarization, $\nu_z = 0$, has the ground state energy $E_0 = 0$.

\subsection{Off-diagonal Form factors}

Away from the chiral limit, the sublattice-bands are not perfectly sublattice polarized, such that the form factors $\Lambda_\bq(\bk)$ are not diagonal in sublattice. We can isolate the off diagonal piece $\Lambda^o$ from the chirally-symmetric diagonal piece $\Lambda^d$
\begin{equation}
  \Lambda_\bq(\bk) = \Lambda_\bq^d(\bk) + \Lambda_\bq^o(\bk), \qquad \Lambda_\bq^{d/o}(\bk) = \frac{1}{2}(\Lambda_\bq(\bk) \pm \sigma_z\Lambda_\bq(\bk) \sigma_z),
  \label{eq:splitformfactor}
\end{equation}
and define the associated density operators
\begin{equation}
  \begin{aligned}
  \delta \hat{\rho}^d_\bq & = \sum_\bk c^\dag_\bk \Lambda^d_\bq(\bk) c_{\bk + \bq} - 4 \ov{\rho}_\bG , \\
  \delta \hat{\rho}^o_\bq & = \hat{\rho}^A_\bq = \sum_\bk c^\dag_\bk \Lambda^o_\bq(\bk) c_{\bk + \bq} 
\end{aligned}
  \label{eq:splitdensities}
\end{equation}
such that the Hamiltonian now reads
\begin{equation}
  \begin{aligned} 
  \H & = \H_d + \H_o \\
  \H_d & = \frac{1}{2A} \sum_\bq V_\bq \delta \hat{\rho}^d_\bq \delta \hat{\rho}^d_{-\bq} \\
  & = \frac{1}{2A} \sum_\bq V_\bq \left( \delta \hat{\rho}^d_\bq \delta \hat{\rho}^o_{-\bq} + \delta \hat{\rho}^o_\bq \delta \hat{\rho}^d_{-\bq} + \delta \hat{\rho}^o_\bq \delta \hat{\rho}^o_{-\bq} \right).
  \end{aligned}
  \label{eq:splitinteraction}
\end{equation}
The Hamiltonian $\H_d$, with densities corresponding to sublattice diagonal form factors, was studied in the previous section. We will now compute the effect of $\H_o$ at first order in perturbation theory. We will use $\langle \cdot \rangle$ to denote an expectation value in one of the ground states of $\H_d$. For the first term of $\H_o$, we have 
\begin{equation}
  \langle \delta \hat{\rho}^d_\bq \delta \hat{\rho}^o_{-\bq} \rangle \propto \langle \delta \hat{\rho}^o_{-\bq} \rangle = \Tr P \rho^o_{-\bq} = \sum_\bk P \Lambda^o_{-\bq}(\bk) = 0
\end{equation}
where we used that ground states of $\H_d$ are eigenstates of $\rho^d$ and that $\Lambda^o$ is off-diagonal in sublattice while $P$ is diagonal. The same argument may be applied to the second term of $\H_o$. 
We therefore have
\begin{equation}
\begin{aligned}
  E_\lambda  & = \langle \H_o \rangle = \frac{1}{2A} \sum_\bq V_\bq \langle \rho^o_\bq \rho^o_{-\bq} \rangle = \frac{1}{2A} \sum_\bq V_\bq \frac{1}{2} \Tr [\rho_\bq^o, P][P,\rho^o_{-\bq}] \\
  & = \frac{1}{2A} \sum_\bq V_\bq \frac{1}{8} \Tr [\rho^o_\bq, Q][Q,\rho^o_{-\bq}]  \\
  & = \frac{1}{16A} \sum_\bq V_\bq \sum_\bk \tr [\Lambda^o_\bq(\bk), Q][Q, \Lambda^{o\dag}_\bq(\bk)] \\
  & = \text{const.} + \frac{1}{8A} \sum_\bq V_\bq \sum_{\bk \tau} \tr  Q^\tau \Lambda^{o \tau}_\bq(\bk) Q^\tau \Lambda^{\dag o \tau}_\bq(\bk). 
  \end{aligned}
  \label{eq:antisotropy}
\end{equation}
In the last step we have expanded the commutators and used $Q^2 = 1$ to write the terms involving $\tr QQ \Lambda^o \Lambda^{o\dag}$  as irrelevant additive constants that we now drop. 

To move forward, we comment on the structure of $\Lambda^o$. Since $\Lambda^o$ is off-diagonal in sublattice and diagonal in valley, we can write $\Lambda^{o \tau}_\bq(\bk) = \sigma_x F^{o \tau}_\bq(\bk) e^{i \Phi^{o \tau}_\bq(\bk) \sigma_z}$ where $\tau$ labels the valley. Time reversal symmetry implies 
\begin{equation}
  \Lambda^{o \tau}_\bq(\bk) = \Lambda^{o (-\tau)}_\bq(-\bk-\bq), \qquad \sum_\bk \left(F^{o \tau}_\bq(\bk) \right)^2 = \sum_\bk \left(F^{o (-\tau)}_\bq(\bk) \right)^2
  \label{eq:trsformfact}
\end{equation}
which we will shortly use.

Using the fact that $Q$ is sublattice and valley diagonal, we have
\begin{equation}
  \begin{aligned}
  E_\lambda & =  \frac{1}{8A} \sum_\bq V_\bq \sum_{\bk \tau} \tr  Q^\tau \Lambda^{o \tau}_\bq(\bk) Q^\tau \Lambda^{\dag o}_\bq(\bk) \\
  & =  \frac{1}{8A} \sum_\bq V_\bq \sum_{\bk \tau} \tr  Q \sigma_x F^{o \tau}_\bq(\bk) Q \sigma_x F^{A \tau}_\bq(\bk) \\
  & =  \frac{1}{8A} \sum_\bq V_\bq \sum_\tau \left[\sum_{\bk} \left(F^{o \tau}_\bq(\bk)\right)^2 \right]  \tr  Q^\tau \sigma_x Q^\tau \sigma_x \\
& =   N_{M}\sum_\tau \frac{\lambda}{4}  \tr  Q^\tau \sigma_x Q^\tau \sigma_x \\
  & =  N_{M} \frac{\lambda}{4} \tr Q \sigma_x Q \sigma_x = N_{M} \frac{\lambda}{2} \tr Q_A Q_B 
\end{aligned}
  \label{eq:lambdaterm}
\end{equation}
where we have defined
\begin{equation}
  \lambda = \frac{1}{2A} \sum_{\bq} V_\bq \frac{1}{N_M} \sum_\bk \left(F^{o\tau_{\bq}}(\bk)\right)^2 = \frac{1}{2A} \sum_{\bq} V_\bq \frac{1}{N_M} \sum_\bk \frac{1}{8} \tr \Lambda^{o  \dag}_\bq(\bk) \Lambda^{o}_\bq(\bk)
  \label{eq:lambdadefn}
\end{equation}
which is independent of the valley $\tau$ due to \eqref{eq:trsformfact}.

\subsection{Superexchange correction from Dispersion}

We now compute the leading correction from the single-particle dispersion $\hat{h}$. We split $h$ into its sublattice diagonal and off-diagonal parts as
\begin{equation}
  h = h_d + h_o, \qquad h_{d,o} = \frac{1}{2} (h + \sigma_z h \sigma_z).
  \label{eq:dispersionsplit}
\end{equation}
The dispersion is diagonal in valley, and the dispersions in each valley are related by time-reversal symmetry 
\begin{equation}
  h^{-\tau}_{o,d}(\bk) = \overline{h^\tau_{o,d}(-\bk)},
  \label{eq:timereversaldispersion}
\end{equation}
where the complex conjugation flips the sign of the part of the off-diagonal dispersion, $h_o = h_x \sigma_x + h_y\sigma_y$, that is proportional to $\sigma_y$. Furthermore particle-hole-inversion symmetry implies
\begin{equation}
  h^{\tau}_{o,d}(\bk) = -h^\tau_{o,d}(-\bk),
  \label{eq:PHInvdispersion}
\end{equation}

The unperturbed states are eigenstates of $h_d = h_0 + h_z \sigma_z$, since all such states correspond to filling bands of definite sublattice and valley,
\begin{equation}
  \begin{aligned}
    \hat{h}_d \ket{\Psi_0} = \left(\sum_\bk \tr h_d(\bk) P \right)  =\tr P \left(\sum_\bk  h_d(\bk)  \right)  \ket{\Psi} = 0
\end{aligned}
  \label{eq:diagonaldispersioneval}
\end{equation}
where we used that $\sum_\bk h_d(\bk) = 0$ due to particle-hole inversion symmetry \eqref{eq:PHInvdispersion}. We will therefore ignore the diagonal dispersion, as it annihilates all unperturbed states.

To split the unperturbed states with dispersion, we must therefore focus on $\hat{h}_o$. Since all unperturbed states are sublattice diagonal, $\langle \hat{h}_0 \rangle = \Tr h_o P = 0$. However, the unperturbed states are not all eigenstates of $\hat{h}_o$, since $\hat{h}_o$ will create particle-hole excitations between the A,B sublattices. For states that have partiallly filled spin and valley flavors, these excitations are not Pauli blocked, and the ability to have such excitations will favor these states within second-order perturbation theory as we shall see.

In order to perform the second-order perturbation theory, we must invert the unperturbed Hamiltonian within the space of excitations and compute
\begin{equation}
  E_2 = -\bra{\Psi_0} \hat{h} (\H_0 - E_0)^{-1} \hat{h} \ket{\Psi_0}
  \label{eq:generalpert}
\end{equation}
While the full space of excitations on top of a state $\ket{\Psi_0}$ is enormous and hard-to control, and inverting $\H$ in this full space would be challenging, we can use the fact that $\hat{h}_o$ only creates particle-hole excitations on top of $\ket{\Psi_0}$; we will soon recall\cite{bultinckGroundStateHidden2020,vafekRenormalizationGroupStudy2020} that the Hamiltonian leaves this block invariant. 
Such excitations, on top of a particular choice of $\ket{\Psi_0}$, form a Hilbert space spanned by
\begin{equation}
  \hat{\phi}\ket{\Psi_0}, \qquad \{ \phi, Q \} = \{ \phi, \sigma_z \} = [\phi, \tau_z] = 0,
  \label{eq:particlehole_hilbertspace_def}
\end{equation}
that is, the creation of particle-hole pairs that are not Pauli Blocked, between sublattices, and within the same valley.

It is straightforward to check that the unperturbed Hamiltonian preserves the Hilbert-space \eqref{eq:particlehole_hilbertspace_def}; to do so we use that $\delta \hat{\rho}^d_\bq \ket{\Psi_0} = \lambda_\bq \ket{\Psi_0}$. For brevity, we will drop the superscript ``$d$'' in this section as we have dealt with the $\Lambda^o$ form factors in the previous section. We have
\begin{equation}
  \begin{aligned}
  \delta \hat{\rho}_\bq \delta \hat{\rho}_{-\bq} \hat{\phi} \ket{\Psi_0} & = \delta \hat{\rho}_\bq\left([\hat{\rho}_{-\bq}, \hat{\phi}] + \hat{\phi} \lambda_{-\bq}\right) \ket{\Psi} \\
  & = \left(\widehat{[\rho_\bq,[\rho_{-\bq},\phi]]} + \lambda_\bq \widehat{[\rho_{-\bq}, \phi]} + \lambda_{-\bq}\widehat{[\rho_{\bq}, \phi]} \right) \ket{\Psi_0}.
\end{aligned}
  \label{eq:Closed_Hilbertspace}
\end{equation}
The operator acting on $\ket{\Psi_0}$ above is a two-body operator that anticommutes with $Q$ if $\phi$ does (note that $[Q,\rho_\bq]=0$: both matrices are diagonal in sublattice and valley, $\rho$ is the identity in spin-space, and $Q$ is $\bk$-independent). Thus, the state \eqref{eq:Closed_Hilbertspace} lies in the Hilbert space spanned by \eqref{eq:particlehole_hilbertspace}. We can therefore invert the Hamiltonian by computing the matrix elements $(H_{\text{eh}})_{\bk' \nu,\bk \mu} = \bra{\Psi_{\bk' \nu}} (\H - E_0) \ket{\Psi_{\bk \mu}}$, where $\ket{\Psi_{\bk \mu}}$ form a basis of particle hole excitations, and then inverting $H_{\text{eh}}$ \cite{bultinckGroundStateHidden2020}. The second order correction then reduces to
\begin{equation}
  E_2 = -\bra{\Psi_0} \hat{h} H_{\text{eh}}^{-1} \hat{h} \ket{\Psi_0}.
  \label{eq:particlehole_hilbertspace}
\end{equation}

Here however we will take a slightly different approach, which is somewhat less algebraically intensive than computing \eqref{eq:particlehole_hilbertspace} directly. In particular we will write 
\begin{equation}
  \ket{\Psi_M} = (1+i\hat{M})\ket{\Psi_0} \approx e^{i \hat{M}} \ket{\Psi_0}, \quad \{M, Q \} = \{M, \sigma_z \} = 0, \quad (M)_{\bk \tau \bk' \tau'} = M^\tau_\bk \delta_{\bk \bk'} \delta_{\tau \tau'},
  \label{eq:variationalperttheory}
\end{equation}
where $\tau$ labels valley, and $M_\bk^\tau$ is a $4 \times 4$ matrix in spin and sublattice space.
The state \eqref{eq:variationalperttheory} should be regarded as a fully-general ansatz for the first-order perturbed wavefunction; $\hat{M}$ should be thought of as $O(\hat{h})$, and it is kept to linear order in the state $\ket{\Psi_M}$ corresponding to the fact that first order perturbations in the wavefunction lead to second order corrections in the energy. The true first order wavefunction corresponds to \eqref{eq:variationalperttheory} with the $M$ that minimizes the total energy 
\begin{equation}
  E_M = \bra{\Psi_M} (\H_0 - E_0 + \hat{h}) \ket{\Psi_M}.
  \label{eq:Menergy_start}
\end{equation}
The simplification that this restructuring leads to is that $\ket{\Psi_M} = e^{i \hat{M}} \ket{\Psi_0}$ is a Slater determinant with $P_M  = e^{-i M} P e^{i M}$. The first order correction does not take us out of the Slater determinant space of states, and we can make use of this to evaluate \eqref{eq:Menergy_start}. 

We proceed to evaluate and minimize \eqref{eq:Menergy_start}. To do the minimization over the space of matrices $M_\bk^\tau$, we will use that they are off diagonal in sublattice
\begin{equation}
  M_\bk^\tau = \begin{pmatrix} 0 & m_\bk^\tau \\ m_\bk^{\tau \dag} & 0 \end{pmatrix}_{A,B}, 
  \label{eq:expandM}
\end{equation}
and expand in a matrix basis
\begin{equation}
m_{\bk}^\tau = \sum_\mu m_{\mu \bk}^\tau r_\tau^\mu, \qquad r^\mu_\tau = - Q^\tau_A r_\tau^\mu Q^\tau_B, \qquad \tr r_\tau^\mu r_\tau^\nu = \delta^{\mu\nu}
  \label{eq:matrixbasis}
\end{equation}
Here, $r^\mu_\tau$ are a basis of $4 \times 4$ matrices that span the subspace of matrices that are odd under the map $m \to Q^\tau_A m Q^\tau_B$ (which squares to the identity).

We begin with the interaction term
\begin{equation}
  \begin{aligned}
    E_{\text{int}}[M] & = \bra{\Psi_M} (\H_0 - E_0) \ket{\Psi_M} = \frac{1}{2A} \sum_\bq V_\bq \langle \delta \hat{\rho}_\bq \delta \hat{\rho}_{-\bq} \rangle_M  - E_0,  \\
    & = \frac{1}{2A} \sum_\bq V_\bq \frac{1}{8}\Tr [Q_M,\rho_\bq][\rho_{-\bq},Q_M] + \frac{1}{2A} \sum_\bq V_\bq \frac{1}{4}\Tr Q_M \rho_\bq \Tr Q_M \rho_{-\bq} - E_0.
    \end{aligned}
  \label{eq:fockhartreeM}
\end{equation}
We must now expand two second order in $M$ via $Q_M = e^{-iM} Q e^{iM} =  e^{i \ad_M} Q = Q + i \ad_M Q  -\frac{1}{2} \ad^2_M Q + \cdots$ where $\ad_M Q = [M,Q] = 2MQ$ since $M$ anticommutes with $Q$. The zeroth order terms reproduce the ground state energy $E_0$. 

\subsubsection{Charge Neutrality Superexchange}

We will begin with the first, Fock, term. We will later see that the Hartree term, the second term in the final line of \eqref{eq:fockhartreeM}, vanishes at charge neutrality. Away from charge neutrality, this term leads to a minor, quantitative, change to the superexchange scale that we compute in the next subsection. For now, however, we will focus on charge neutrality and the Fock term.

In the Fock term, we must expand each of $[Q_M, \rho_\bq]$ to at least linear order because $[Q, \rho_\bq]=0$. We therefore have, to second order in $M$,
\begin{equation}
  \begin{aligned}
    \frac{1}{2A} \sum_\bq V_\bq \frac{1}{8}\Tr [Q_M,\rho_\bq][\rho_{-\bq},Q_M] & = \frac{1}{2A} \sum_\bq V_\bq (-)\frac{1}{8}\Tr [[M,Q],\rho_\bq][\rho_{-\bq},[M,Q]],  \\
    &  = \frac{1}{2A} \sum_\bq V_\bq \left(\frac{1}{4}\Tr MM(\rho_\bq\rho_{-\bq} + \rho_{-\bq} \rho_{\bq}) - \frac{1}{2} \Tr M \rho_\bq M \rho_{-bq}\right), \\
    & = \frac{1}{4A} \sum_\bq V_\bq \sum_\bk \tr \left(M_\bk^2 \Lambda_\bq(\bk) \Lambda_{-\bq}(\bk) - M_\bk \Lambda_\bq(\bk) M_{\bk + \bq} \Lambda_{-\bq}(\bk+\bq) \right), \\
  & = \frac{1}{4A} \sum_\bq V_\bq \sum_{\bk \tau} \bigg( \tr(m^{\tau\dag}_\bk m_\bk^\tau) \left(\Lambda^{A\tau}_\bq(\bk) \Lambda^{A \tau}_{-\bq}(\bk+\bq) + \Lambda^{B\tau}_\bq(\bk) \Lambda^{B \tau}_{-\bq}(\bk+\bq)\right)    \\
  & \qquad \qquad - \tr(m^{\tau \dag}_\bk m^\tau_{\bk + \bq}) \Lambda^{A\tau}_\bq(\bk) \Lambda^{B \tau}_{-\bq}(\bk+\bq) -  \tr(m^{\tau \dag}_{\bk + \bq} m^\tau_{\bk}) \Lambda^{A \tau}_{-\bq}(\bk+\bq) \Lambda^{B\tau}_\bq(\bk) \bigg), \\
& = \frac{1}{2A} \sum_\bq V_\bq \sum_{\bk \tau} \tr(m^{\tau\dag}_\bk m_\bk^\tau) \frac{1}{2}\left(\Lambda^{A\tau}_\bq(\bk) \Lambda^{A \tau}_{-\bq}(\bk) + \Lambda^{B\tau}_\bq(\bk) \Lambda^{B \tau}_{-\bq}(\bk)\right) \\
& \qquad \qquad \qquad -\tr(m^{\tau \dag}_\bk m^\tau_{\bk + \bq})\Lambda^{A\tau}_\bq(\bk) \Lambda^{B \tau}_{-\bq}(\bk+\bq) )\\
& = \frac{1}{2} \sum_{\bk, \bk', \tau} \overline{m}_{\mu\bk}^\tau R^{F, \tau}_{\bk \bk'} m_{\mu \bk'}^\tau,
\end{aligned}
  \label{eq:massagefock}
\end{equation}
where
\begin{equation}
  R^{F, \tau}_{\bk \bk'} = \frac{1}{A} \sum_\bq V_\bq\left( \frac{1}{2} \delta_{\bk \bk'}(\abs{\Lambda^A_\bq(\bk)}^2 + \abs{\Lambda^B_\bq(\bk)}^2) - 
  \Lambda^{A\tau}_\bq(\bk) \Lambda^{B \tau}_{-\bq}(\bk+\bq) \delta_{\bk', \bk+\bq} \right),
  \label{eq:fockenergy}
\end{equation}
is a Fock Hamiltonian that describes the exchange energy penalty for creating particle-hole pairs in valley $\tau$. In simplifying \eqref{eq:massagefock}, we repeatedly used $[Q, \rho_\bq] = \{M ,Q \} = 0$, the cyclic property of the trace, and $V(\bq) = V(-\bq)$. The Fock Hamiltonians in the two valleys are related by time reversal symmetry. Indeed, using that the form factors satisfy $\Lambda_\bq^\tau(\bk) = \ov{\Lambda_{-\bq}(-\bk)}$ we have
\begin{equation}
  R^{\tau}_{F \bk \bk'} = \overline{R}^{-\tau}_{F-\bk, -\bk'} = R^{\tau}_{F-\bk',-\bk} = \left(T \left(R_F^{(-\tau)}\right)^T T\right)_{\bk \bk'},
  \label{eq:transposeTRS}
\end{equation}
where $T_{\bk \bk'} = \delta_{\bk,-\bk'}$ is the matrix that takes $\bk \to -\bk$ and we have written the action of time reversal ultimately as a transpose, related to complex conjugation by the, verifiable, hermiticity of $R$.

We now compute the dispersion term, which contributes at first order in $M$ and thus favors a nonzero perturbative correction $M$ of the appropriate direction and sign. We will write
\begin{equation}
  h_{o \tau \bk} = \begin{pmatrix} 0 & z_{\tau \bk} \\ \ov{z}_{\tau \bk} & 0 \end{pmatrix}
  \label{eq:offdiagonaldispersionform}
\end{equation}
such that
\begin{equation}
  \langle \hat{h}_o \rangle_M = \frac{i}{2}\Tr h_o [M,Q] = -i \Tr QM h_o = 
  -i \sum_{\mu \tau \bk}\left( m_{\mu \bk}^\tau  \ov{z}_{\tau \bk} \tr Q_A^\tau r^\mu_\tau - \ov{m}_{\mu \bk}^\tau z_{\tau \bk} \tr Q_A^\tau r^{\mu \dag}_\tau \right).
  \label{eq:dispersionterm}
\end{equation}

In total, the energy as a function of $m_\mu$ is 
\begin{equation}
  E[M]  =  \frac{1}{2} \sum_{\bk, \bk', \tau} \overline{m}_{\mu\bk}^\tau R^{F, \tau}_{\bk \bk'} m_{\mu \bk'}^\tau,
  -i \sum_{\mu \tau \bk}\left( m_{\mu \bk}^\tau  \ov{z}_{\tau \bk} \tr Q_A^\tau r^\mu_\tau - \ov{m}_{\mu \bk}^\tau z_{\tau \bk} \tr Q_A^\tau r^{\mu \dag}_\tau \right).
  \label{eq:totalenergy_M}
\end{equation}
As discussed above, the energy in second order perturbation theory corresponds to the minimum of \eqref{eq:totalenergy_M} over all $m_\mu$ because our variational space encompasses all possible perturbative corrections. The minimum energy is then
\begin{equation}
  E_{\text{SE}} = \min_M E[M] =   -2 \sum_\tau \sum_\mu\tr Q_A^\tau r^\mu_\tau \tr Q_A^\tau r^{\dag\mu}_\tau \sum_{\bk \bk'} \overline{z}_{\bk \tau} (R_F^\tau)^{-1}_{\bk \bk'} z_{\bk \tau}.
  \label{eq:SEenergy_prefierze}
\end{equation}
We now define
\begin{equation}
  J_{\nu = 0} = \frac{2}{ N_{M}}\sum_{\bk \bk'}  \overline{z}_{\bk \tau} (R_F^\tau)^{-1}_{\bk \bk'} z_{\bk \tau}
  \label{eq:Jdefn}
\end{equation}
as the superexchange energy per unit cell at charge neutrality, where $N_{M}$ is the number of unit cells and number of $\bk$-points. While \eqref{eq:Jdefn} has $\tau$ dependence, in fact $J$ is independent of the valley $\tau$ due to the time reversal symmetries \eqref{eq:transposeTRS} and \eqref{eq:timereversaldispersion}, the latter of which implies $z_{\bk \tau} = \overline{z}_{-\bk -\tau} = (T \overline{z}_{-\tau})_\bk$. 

We must now deal with the sum over generators $r^\mu_\tau$ using an appropriate Fierz identity for the space of matrices odd under $m \mapsto -Q^\tau_A m Q^\tau_B$. This identity can be derived from the usual Fierz identity of all matrices, $\tr X t^\mu \tr Y t{\dag \mu}$ together with the projection $t^\mu \to -Q^\tau_A t^\mu Q^\tau_B = r^\mu$:
\begin{equation}
  \sum_\mu \tr X r^\mu \tr Y r^{\dag \mu}_\tau = \frac{1}{2}(\tr X Y - \tr Q_B X Q_A Y).
  \label{eq:fierze}
\end{equation}
Substituting $X = Y = Q_A$ we have
\begin{equation}
  E_{\text{SE}} = \frac{1}{2} J_{\nu = 0} \sum_\tau (\tr Q^\tau_A Q^\tau_B - 2) = \frac{1}{2} J_{\nu = 0}  (\tr Q_A Q_B - 4)
  \label{eq:CNPfinal}
\end{equation}
as the superexchange energy at charge neutrality

\subsubsection{Superexchange Away from Charge Neutrality}

We now perform a full calculation that includes the Hartree term and is valid at all integer fillings. We must compute the Hartree energy, the second term in the second line of \eqref{eq:fockhartreeM}. We use that $\Tr [M,Q] \rho_\bq = 0$ since $M$ is off-diagonal in sublattice while $Q$ and $\rho_\bq$ are diagonal. Then, the terms contributing to second order are, again using $V(\bq) = V(-\bq)$,
\begin{equation}
  \begin{aligned}
    E_H[M] = \frac{1}{2A} \sum_\bq V_\bq \frac{1}{4}\Tr Q_M \rho_\bq \Tr Q_M \rho_{-\bq} & = -\frac{1}{2A} \sum_\bq V_\bq \frac{1}{4}\Tr [M,[M,Q]] \rho_\bq \Tr Q \rho_{-\bq} \\
  & = -\frac{1}{2A} \sum_\bG V_\bG \frac{1}{2} \sum_{\bk \tau} \left(\Lambda^{0 \tau}_\bG(\bk) \tr([M^\tau_\bk,[M^\tau_\bk,Q^\tau]]) + \Lambda^{z \tau}_\bG(\bk) \tr([M^\tau_\bk,[M^\tau_\bk,Q]]\sigma_z ) \right) \\
  & \qquad \qquad \times \left(\nu \overline{\rho}_{-\bG} + \nu_z \overline{\rho}^z_{-\bG} \right) \\
& = -\frac{1}{2A} \sum_\bG V_\bG \frac{1}{2}\sum_{\bk \tau} \Lambda^{z \tau}_\bG(\bk) \tr([M^\tau_\bk,[M^\tau_\bk,Q]]\sigma_z)  \left(\nu \overline{\rho}_\bG + \nu_z \overline{\rho}^z_{\bG} \right), \\
& = -\frac{1}{A} \sum_\bG V_\bG \sum_{\bk \tau} \Lambda^{z \tau}_\bG(\bk) \tr(M^\tau_\bk M^\tau_\bk Q^\tau \sigma_z)  \left(\nu \overline{\rho}_\bG + \nu_z \overline{\rho}^z_{\bG} \right), \\
& = -\frac{1}{A} \sum_\bG V_\bG \sum_{\bk \tau} \left( \tr m^\tau_\bk m^{\tau \dag}_\bk Q_A^\tau - \tr m^{\tau \dag}_\bk m^\tau_\bk Q_B^\tau \right) \Lambda^{z \tau}_\bG(\bk) \left(\nu \overline{\rho}_\bG + \nu_z \overline{\rho}^z_{\bG} \right), \\
    \end{aligned}
  \label{eq:HartreeTerm}
\end{equation}
where we decomposed $Q$ into valleys as $Q = \sum_\tau Q^\tau \frac{1 + \tau \tau_z}{2}$. We note that at charge neutrality the ground states have zero sublattice polarization, such that $\nu = \nu_z = 0$ and the term \eqref{eq:HartreeTerm} can be dropped. Furthermore, in more symmetric systems like twisted bilayer graphene $\Lambda^z_\bG(\bk) = 0$, so that again \eqref{eq:HartreeTerm} can be dropped. However, for h-HTG, away from charge neutrality, we must include \eqref{eq:HartreeTerm}. 

We note that due to the presence of $Q_A$ and $Q_B$ in the traces in the final line, the Hartree contribution to the particle-hole Hamiltonian is not the identity matrix in the space of generators $r^\mu$; in contrast the Fock Hamiltonian \eqref{eq:massagefock}, \eqref{eq:fockenergy} did not depend on $\mu$. However, we can carefully choose a basis in which the Hartree energy is diagonal in the space of generators such that inversion is still straightforward. To do this, we split the problem into symmetry unrelated cases based on $\frac{1}{2}\tr Q_{A,B}^\tau = 0, \pm 1$, which are invariant under the $(U(2))^4$ unperturbed symmetry of sublattice and valley resolved spin and charge rotations. We will later merge the final results of this case study with a formula that encompasses all cases.

We begin with the cases $\frac{1}{2} \tr Q_A^\tau = \zeta$ or $\frac{1}{2}\tr Q_B^\tau = -\zeta$, where $\zeta = \pm 1$, such that $Q^\tau_A = \zeta$ or $Q^\tau_B = -\zeta$ is proportional to the $2 \times 2$ identity matrix in spin-space. We note that $\tr m^\tau_\bk m^{\tau \dag}_\bk Q_A^\tau = -\tr m^{\tau \dag}_\bk m^\tau_\bk Q_B^\tau$, so that we can replace $\tr m^\tau_\bk m^{\tau \dag}_\bk Q_A^\tau - \tr m^{\tau \dag}_\bk m^\tau_\bk Q_B^\tau$ with $2 \zeta \tr m^\dag m$ by converting the $Q^\tau_B$ term to the $Q^\tau_A$ one if $Q^\tau_A = \zeta$, or vice versa if $Q^\tau_B = -\zeta$. We note that we have implicitly excluded the case $Q^\tau_A = Q^\tau_B = \pm 1$, but this is because there are no particle-hole excitations in valley $\tau$ in this case: the condition $m = -Q_A m Q_B$ reduces to $m = -m$. We will see that this $Q_A = Q_B$ case is dealt with appropriately in that the superexchange energy will vanish.

The Hartree energy for $Q_A = \zeta$ or $Q_B = -\zeta$, for $\zeta = \pm 1$ and in valley $\tau$, is then
\begin{equation}
  E_{H \tau}[M] = \zeta \sum_{ \bk \mu} \ov{m}^\tau_{\mu \bk} R^\tau_{H \bk, \bk'} m^\tau_{\mu \bk'}, \qquad R^\tau_{H \bk \bk'} = - \delta_{\bk \bk'}\frac{2}{A} \sum_\bG V_\bG \Lambda^{z \tau}_\bG(\bk) \left(\nu \overline{\rho}_\bG + \nu_z \overline{\rho}^z_{\bG} \right).
  \label{eq:hartree_energy_traceful}
\end{equation}
To obtain the corresponding superexchange energy associated with valley $\tau$, we follow the previous subsection with $R^\tau_F \to R^\tau_F + \zeta R^\tau_H$. We then have
\begin{equation}
  E^\tau_{\text{SE}} = \frac{1}{2}J_\zeta (\tr Q_A^\tau Q_B^\tau - 2) = \frac{1}{2}J_\zeta (\tr Q_A^\tau Q_B^\tau - 2) = \frac{1}{2}\zeta J_\zeta (\tr Q_B^\tau - \tr Q_A^\tau)
  \label{eq:superenergy_traceful}
\end{equation}
where
\begin{equation}
  J_{\zeta} = \frac{1}{2 N_{M}}\sum_{\bk \bk'}  \overline{z}_{\bk \tau} (R_F^\tau + \zeta R_H^\tau)^{-1}_{\bk \bk'} z_{\bk \tau}
  \label{eq:Jdefn_2}
\end{equation}
does not depend on $\tau$ due to time reversal symmetry.

We now deal with the other case, where $\tr Q_A^\tau = \tr Q_B^\tau = 0$. While the particle-hole Hamiltonian is no longer the identity matrix in the space of generators, we will choose a basis so that it is (block) diagonal. In particular we choose $r^\mu$ such that $Q_A^\tau r^\mu = -r^\mu Q_B^\tau = \zeta_\mu r^\mu$ where $\zeta_\mu = (+1,-1)$. Such a basis always exists; to see this we diagonalize $Q^\tau_A = U^\tau_A s_z U^{\tau \dag}_A$ and $Q^\tau_B = U^\tau_B (-s_z) U^{\tau \dag}_B$ independently, with two different unitaries $U^\tau_{A,B}$, and define $r^\mu = U^\tau_A \frac{1 \pm s_z}{2} U^\dag_B$. The Hartree energy is, in this case,
\begin{equation}
  E_{H, \tau} = \sum_{\bk \bk' \mu} \zeta_\mu \ov{m}^\tau_{\mu \bk} R^\tau_{H \bk, \bk'} m^\tau_{\mu \bk'},
  \label{eq:hartreeenergy_zerotr}
\end{equation}
so that the total interacting particle-hole Hamiltonian is diagonal in this basis
\begin{equation}
  R^{\tau \mu \nu}_{\bk \bk'} = (R^\tau_{F \bk \bk'} + \zeta_\mu R^\tau_{H \bk \bk'}) \delta^{\mu \nu}
  \label{eq:diagonal_notidentity}
\end{equation}
and straightforward to invert in each block separately. The resulting superexchange energy in valley $\tau$ is
\begin{equation}
  E^\tau_{\text{SE}} =  -\sum_\mu J_{\zeta_\mu} \tr Q_A^\tau r^\mu_\tau \tr Q_A^\tau r^{\dag\mu}_\tau  =  -J \sum_\mu \tr Q_A^\tau r^\mu_\tau \tr Q_A^\tau r^{\dag\mu}_\tau -  J^{(-)} \sum_\mu \zeta_\mu \tr Q_A^\tau r^\mu_\tau \tr Q_A^\tau r^{\dag\mu}_\tau,
\end{equation}
where we defined $J^{(\pm)} = \frac{1}{2}(J_{\zeta = +} \pm J_{\zeta = -})$ and $J = J^{(+)}$. We now use $Q_A^\tau r^\mu = \zeta_\mu r^\mu$ in our choice of basis in one of the traces in the second term to get rid of the $\zeta_\mu$ pre-factor which lets us apply the Fierz identity \eqref{eq:fierze} to both terms:
\begin{equation}
  E^\tau_{\text{SE}} = -J \sum_\mu \tr Q_A^\tau r^\mu_\tau \tr Q_A^\tau r^{\dag\mu}_\tau - J^{(-)} \sum_\mu  \tr r^\mu_\tau \tr Q_A^\tau r^{\dag\mu}_\tau = \frac{1}{2}J (\tr Q_A^\tau Q_B^\tau -2)  - \frac{1}{2} J^{(-)} \left( \tr Q_A^\tau - \tr Q_B^\tau \right).
  \label{eq:superenergy_zerotrace}
\end{equation}
While the second term above is zero under our assumption of $\tr Q_A = \tr Q_B = 0$, we note that if we keep it then the above expression also reproduces \eqref{eq:superenergy_traceful} for $\tr Q_A = \zeta$ or $\tr Q_B = -\zeta$. Therefore, \eqref{eq:superenergy_zerotrace} actually encompasses both cases, and we can cease our case study. Combining both valleys we have
\begin{equation}
  E^\tau_{\text{SE}} = \frac{1}{2} J \tr Q_A Q_B, \qquad J = \frac{2}{N_{M}} \sum_{\bk \bk'} \overline{z}_{\bk \tau} \frac{1}{2}\left( (R^\tau_F + R^\tau_H)^{-1} + (R^\tau_F - R^\tau_H)^{-1} \right)_{\bk \bk'} z_{\bk' \tau}
  \label{eq:finalanswerfinally}
\end{equation}
up to an additive constant that depends on the sublattice polarization $\nu_z$, but doesn't split the states for a fixed sublattice polarization. We see that we reproduce the answer at charge neutrality, where $R_H$ is zero, and away from charge neutrality there is just a quantitative modification to the overall coefficient.

\section{Representative $S_z$ conserving states}\label{app:spinrot}
While generic strong-coupling states need not satisfy spin $S_z$ conservation, $[Q,S_z]=0$, we argue here that all ground states are symmetry related to a representative state with conserved $S_z$.
To see this, we note that $(J-\lambda) \tr Q_A Q_B = (J-\lambda) \sum_\tau \tr Q_A^\tau Q_B^\tau$ where $Q_A^\tau$ is a $2 \times 2$ matrix that describes the spin occupation and direction associated with sublattice $A$ and valley $\tau$. If $\tr Q_A^\tau = \tr Q_B^\tau = 0$, such that one spin projection in each valley-sublattice sector is filled, then we can write $Q_{\tilde{\sigma}}^\tau = \v{n}_{\tilde{\sigma}}^\tau \cdot \v s$ where $\v{n}_{\tilde{\sigma}}^\tau$ is the direction of the spin in sublattice $\tilde{\sigma}$ and valley $\tau$, and $\v s = (s_x,s_y,s_z)$ are spin Pauli matrices. Then, the 
energetics favor the spin direction of the $A$ sublattice band to be either aligned, for $\lambda > J$, or antialigned, for $J<\lambda$, with the $B$ sublattice for each valley $\tau$, such that both spins lie along the same axis regardless. We can then use the spin rotation symmetry in valley $\tau$ to align this axis with the $z$ axis, such that we arrive at a state with conserved $S_z$ as claimed. If one or both sublattices are fully filled or fully empty then the same conclusion straightforwardly holds.

\section{Flavor permutation symmetry}\label{app:flavor_permutation}
As discussed in the main text, Slater determinant states of h-HTG that are $S^z$ and valley conserving have a ``flavor permutation symmetry'', which we elaborate on here. We will not assume that $P$ describes a strong coupling state. Using the notation of Appendix \ref{app:strong_coupling}, the Hartree Fock energy is
\begin{equation}
    E[P] = \frac{1}{2A} \sum_{\bq} V_\bq \langle \delta \rho_\bq \delta \rho_{- \bq} \rangle = \frac{1}{2A} \sum_\bq V_\bq  \Tr \left( [\rho_\bq, P][P,\rho_{-\bq}] \right) + \frac{1}{2A} \sum_\bG V_\bG \langle \delta \rho_\bG \rangle \langle \delta \rho_{-\bG} \rangle
\end{equation}
The first term is the Fock term and the second term is the Hartree term.  We begin with the Fock term. Since the Fock term has a single trace, and both $\rho_\bq$ and $P$ are diagonal in flavor, we have
\begin{equation}
E_F[P] = \sum_{\tau s}E_{\tau}[P^{\tau s}], \qquad E_{\tau}[P^{\tau s}] = \frac{1}{2A} \sum_\bq V_\bq \frac{1}{2} \Tr \left( [\rho^{\tau s}_\bq, P^{\tau s}][P^{\tau s},\rho^{\tau s}_{-\bq}] \right)
\end{equation}
where the exchange functional $E_{\tau}$ depends on valley but not spin, because of spin rotation symmetry, and $P^{\tau s}$ and $\rho^{\tau s}$ are the $2N_M \times 2N_M$-sized blocks associated with valley $\tau$ and spin $s$ of $P$ and $\rho_\bq$ respectively (note that $\rho^{\tau s}$ is $s$-independent; we kept the label to indicate that its size corresponds to that of a single flavor block). The factor of $2$ in $2N_M$ comes from sublattice and $N_M$ is the number of moir\'{e} cells and the number $\bk$-points in the mBZ. 

We now demonstrate explicitly the flavor permutation symmetry of the Fock term. It is straightforward to see that we can exchange $P^{\tau \uparrow} \leftrightarrow P^{\tau \downarrow}$, for some fixed $\tau$, without changing the Fock energy, since the functional $E_\tau$ does not depend on $s$. This processes is simply making use of the spin rotation symmetry $SU(2)_\tau$. More interestingly, we can use TRS to exchange valleys \emph{within a single, fixed, spin species}. Indeed, time reversal symmetry implies that $\rho_\bq^{\tau} = T\overline{\rho_{-\bq}^\tau } T^{-1}$, where $T$ takes $\bk \to -\bk$ and $\rho_{-\bq} = \rho_\bq^\dag$, so that $E_{\tau}[P^{\tau s}] = E_{{-\tau}}[T \overline{P^{\tau s}} T^{-1}]$. To understand the flavor permutation symmetry involving opposite valleys that the time reversal relation implies, let us begin with a QSH insulator at neutrality, $P_{\text{QSH}}$, consisting of filling both sublattices of the flavors $(K,\uparrow)$, $(\bar{K},\downarrow)$ and leaving the other two flavors empty. By exchanging the flavors $(\bar{K},\downarrow) \leftrightarrow (K, \downarrow)$, through defining 
\begin{equation}
\begin{aligned}
    P_{\text{CI}}^{\tau \uparrow} & =  P^{(-\tau) \uparrow}_{\text{QSH}}  \\
    P_{\text{CI}}^{\tau \downarrow} & = T \overline{P^{(-\tau) \downarrow}_{\text{QSH}}} T^{-1}
    \end{aligned}
    \label{eq:singlespeciesTRS}
\end{equation}
we arrive at a valley polarized $C = 2$ Chern insulator (CI) with the exact same Fock energy. By combining the action of time reversal on individual spin species and spin flip symmetries within each valley, we can permute the occupations of all flavors without changing the Fock energy.

The Hartree energy is also flavor permutation symmetric, as while the Hartree energy contains a product of traces, from the product of the density expectation values $\langle \delta \rho_\bG \rangle \langle \delta \rho_\bG \rangle$ and $\langle \delta \rho_\bG \rangle = \Tr P \rho_\bG - 4 \ov{\rho}_\bG = \Tr Q \rho_\bG$, each trace on its own decomposes into a sum over flavors that is flavor permutation symmetric.
Indeed, TRS implies $\rho_\bG^\tau = T\ov{\rho_{-\bG}^{-\tau}}T^{-1} $ where we used that the form factors satisfy $\Lambda_\bq(\bk + \bG) = \Lambda_\bq(\bk)$ in our periodic gauge $c^\dag_\bk = c^\dag_{\bk + \bG}$. We therefore have that $\Tr Q^{\tau s} \rho^{\tau}_\bG = \Tr T \ov{Q^{\tau s}} T^{-1} \rho^{-\tau}_\bG$, so that the valleys can be permuted in the same way as in the Fock energy.
\end{appendix}

\end{document}